\documentclass{report}
\pdfoutput=1
\usepackage{cmap}
\usepackage[utf8]{inputenc}
\usepackage[T1]{fontenc}

\usepackage[margin=3cm,includehead,includefoot,heightrounded]{geometry}
\usepackage[style=numeric,sorting=none,maxbibnames=99,backref]{biblatex}
\usepackage{xcolor}
\usepackage{xurl}
\PassOptionsToPackage{hyphens}{url}\usepackage{hyperref}
\usepackage{enumitem}
\usepackage{tabularx}
\usepackage{multirow}
\usepackage{graphicx}
\usepackage{csquotes}
\usepackage{subcaption}
\usepackage{verbatim}
\usepackage{caption}
\usepackage[australian]{babel}
\usepackage{doi}
\usepackage{hyphenat}
\usepackage{authblk}

\graphicspath{ {./figures/} }
\title{Distributed Edge-based Video Analytics on the Move}
\author{Jayden King}
\author{Young Choon Lee}
\affil{The School of Computing, Macquarie University, NSW Australia 2109\\
{\tt\{\href{mailto:jayden.king@mq.edu.au}{jayden.king},\href{mailto:young.lee@mq.edu.au}{young.lee}\}@mq.edu.au\\}}
\date{}

\hypersetup{
    colorlinks,
    linkcolor={black},
    citecolor={blue!50!black},
    urlcolor={black}
}

\begin{document}
\maketitle

\widowpenalty=10000
\clubpenalty=10000

\begin{abstract}
    In recent years, dash cams have gained international popularity for personal and commercial use~\cite{mehrishEgocentricAnalysisDashCam2020,parkMotivesConcernsDashcam2016}. Although dash cams are primarily used to collect evidence for traffic incidents, further value may be gained from the videos they record through video analytics. Commercial dash cams lack the resources necessary to perform video analytics, so their video data must be offloaded elsewhere to be processed. Cloud computing is a popular choice for offloading computationally intensive tasks, though the high latency and bandwidth usage of cloud computing is undesirable.

    These issues can be mitigated through edge computing, where processing occurs close to the data source. A device that is likely to be in close proximity to a dash cam is a mobile device, one belonging to either the vehicle's driver or passengers. Modern mobile devices such as smartphones are much more powerful than commercial dash cams, yet they still have a fraction of the resources available to cloud servers. A single smartphone is capable of performing video analytics on dash cam recordings, but may be unable to produce results in a real-time manner. Instead of using a single mobile device, multiple can form a local network to share their resources and perform computationally intensive tasks in a shorter amount of time. With a local network of mobile devices, video analytics can be performed on dash cam recordings while avoiding the disadvantages of cloud computing.

    In this thesis, we present EdgeDashAnalytics (EDA), an edge-based system that enables near real-time video analytics using a network of mobile devices. In particular, it simultaneously processes videos produced by two dash cams of different angles with one or more mobile devices on the move in a near real-time manner. One camera faces outward to capture the view in front of the vehicle, while the other camera faces inward to capture the driver. The outer videos are analysed to detect potential driving hazards, while the inner videos are used to identify driver distractedness.
    It was found that it was not possible to achieve real-time results simply by distributing processing across a local network of mobile devices. Shortcomings of the OS and libraries introduced delays that could be dismissed in other tasks, but cannot be ignored in time sensitive tasks such as video analytics. We have overcome these shortcomings by devising several optimisations. By incorporating these optimisations, EDA achieves near real-time video analytics, mitigating the effect of such delays with a tolerable loss in accuracy. We have implemented EDA as an Android app and evaluated it using two dash cams and several heterogeneous mobile devices with the BDD100K dash cam video dataset~\cite{yuBDD100KDiverseDriving2020} and the DMD driver monitoring dataset~\cite{ortegaDMDLargeScaleMultimodal2020}. Experiment results demonstrate the feasibility of real-time video analytics in terms of turnaround time and energy consumption (or battery usage), using resource-constrained mobile devices on the move.
\end{abstract}

\hypersetup{linkcolor={red!50!black},urlcolor={blue!80!black}}

\chapter{Introduction}\label{chap:introduction}
Commercial dash cams serve an important role in collecting evidence for traffic incidents. Yet the vast majority of video data produced by dash cams do not capture traffic incidents, so they are discarded. Instead of wasting this video data, further value may be gained from them with video analytics. As the name suggests, video analytics concerns the automated analysis of video data, typically with the use of machine learning techniques. Examples of video analytics includes tasks such as object detection, object tracking, and facial recognition. Commercial dash cams are dedicated recording devices, they lack the resources to perform these computationally intensive tasks. Therefore, in order to perform video analytics on the video data produced by dash cams, said video data must be offloaded to a device with greater processing capacity. Cloud computing is a popular choice for offloading computationally intensive tasks as it is relatively cheap and easy to use. However, as there is typically a great distance between cloud servers and data sources like dash cams, the latency involved in data transmission is unsuitable for time-sensitive tasks such as live video analytics. Additionally, sending all of the video data produced by dash cams to could servers would consume an enormous amount of bandwidth. This would strain network infrastructure and may create significant problems for those with data-limited internet plans.
These issues could be avoided with edge computing, an alternative to cloud computing where data is only transmitted a short distance. Instead of sending video data to distant cloud servers, video data may be sent to nearby mobile devices such as smartphones. Although smartphones have much greater processing capacities than commercial dash cams, it is unlikely that a single smartphone would be capable of processing the video data produced by two cameras in a real-time manner. However, this could become possible through optimisations and distributing the video data across a local network of smartphones for processing.

There are a number of challenges in using a mobile device to perform video analytics on dash cam video. Firstly, commercial dash cams are typically provided with a closed-source companion app instead of offering a public API. This is a significant complication, as it prevents direct programmatic interaction between dash cams and mobile devices. Instead of being able to access a live video stream from a dash cam, a mobile device can only use a HTML interface to download recorded video files. One dash cam did provide a live stream accessible by mobile devices, but this live stream was disabled whenever the dash cam's secondary camera was connected. This means that video files are the only type of data accessible on commercial dash cams, leading to the next issue. There are no video analytics or machine learning libraries available on mobile platforms that directly support video files, at best they may support the use of the device's own camera feed. These libraries instead accept bitmap data, meaning that in order to use video files with these libraries, the video's frames must first be extracted as bitmaps. Frame extraction is a relatively slow process, however, the fastest method was found to be provided by the Android standard library. Despite this, frame extraction was slow enough to prevent real-time turnaround. Only by optimising the other processes involved in video analysis was it possible to reach near real-time turnaround. Another issue is that of video granularity, the length of video files. While it would be desirable to have the smallest possible granularity, shortcomings inherent to mobile systems make it infeasible to use granularities below a certain point. This is due to the overhead delays involved in handling individual video files, delays that do not directly scale with the video file's length. It was determined that two seconds was the smallest granularity possible when downloading videos from a dash cam, while one second was possible when downloads were simulated.

In this research, we present EdgeDashAnalytics (EDA)\footnote{EDA GitHub repository: \url{https://github.com/JaydenKing32/EdgeDashAnalytics}}, an edge-based system that utilises a network of mobile devices to perform video analytics on dash cam video in a near real-time manner. The system analyses the video data from both outward and inward facing cameras in order to improve the safety of driving. We have devised several optimisations and incorporated them into EDA to overcomes the challenges described above. These optimisations include simultaneous video download, scheduling, segmentation and early stopping. By incorporating these optimisations, EDA achieves near real-time video analytics, mitigating the effect of such delays with a tolerable loss in accuracy. We have implemented EDA as an Android app and evaluated it using two dash cams and several heterogeneous mobile devices with the BDD100K dash cam video dataset~\cite{yuBDD100KDiverseDriving2020} and the DMD driver monitoring dataset~\cite{ortegaDMDLargeScaleMultimodal2020}. Experiment results demonstrate the feasibility and potential of real-time video analytics in terms of turnaround time and energy consumption (or battery usage), using resource-constrained mobile devices on the move.

The case study chosen to demonstrate EDA involves the simultaneous analysis of video data produced by two different cameras. Many commercial dash cams include a secondary camera intended to record the rear of the vehicle. However, we repurposed the secondary camera to record the driver instead. Each camera is associated with a separate video analytics task. The task associated with the outward-facing camera is to detect potential road hazards and identify if the driver is tailgating. The task associated with the inward-facing camera is to detect driver distractedness, in other words, identifying actions performed by the driver that are unrelated to operating the vehicle. This aims to improve driving safety in two ways. The primary concern is avoiding road hazards that may damage the vehicle or even cause a crash. The secondary concern is preventing driver distractedness, as focusing on anything besides driving can greatly increase the likelihood of an accident~\cite{dingusDriverCrashRisk2016}.

\begin{figure}[htb]
    \centering
    \includegraphics[width=\textwidth]{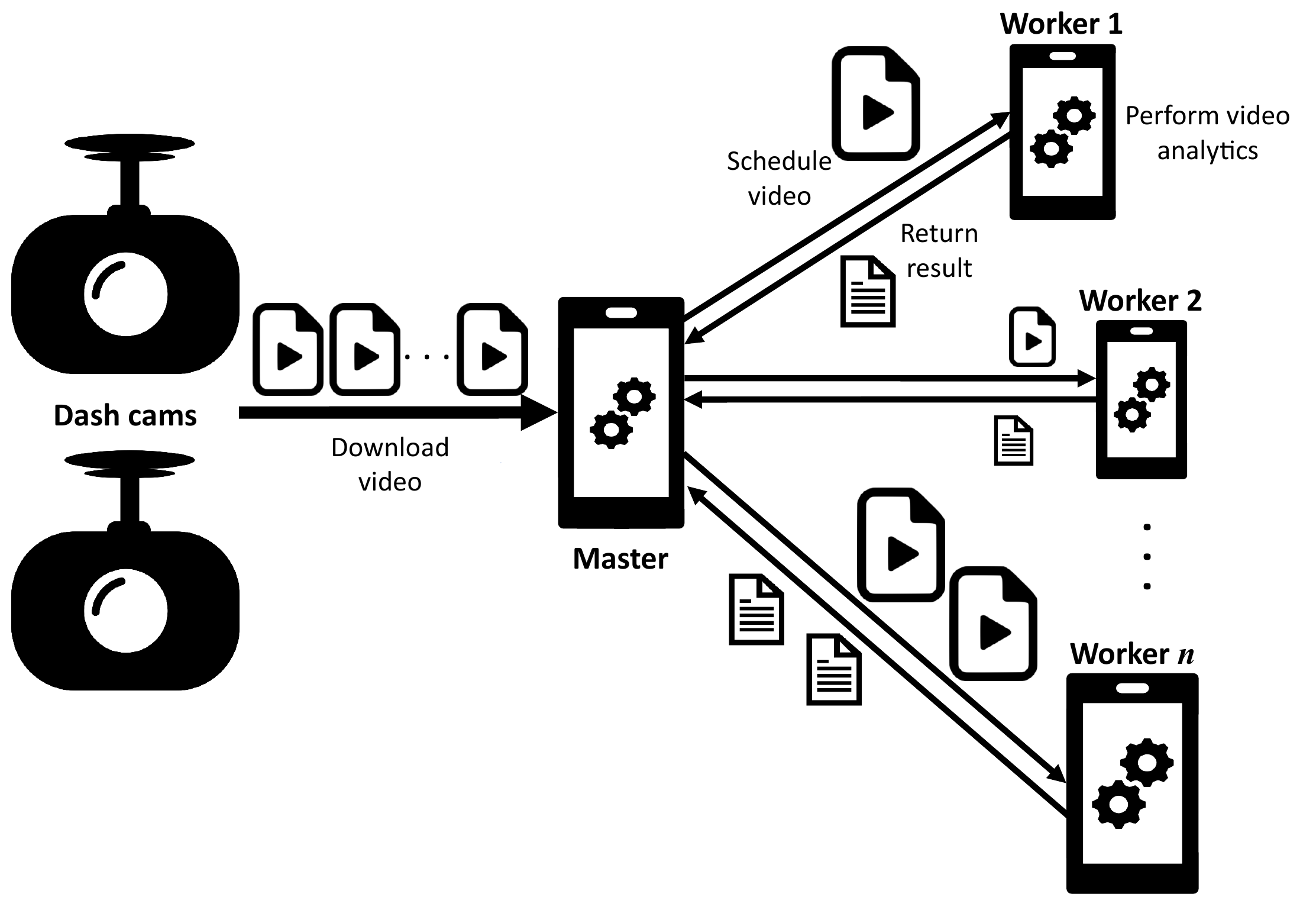}
    \caption{Brief overview of EDA.}
    \label{fig:overview_brief}
\end{figure}

Figure~\ref{fig:overview_brief} demonstrates a brief overview of EDA. It shows a master device downloading video files from two dash cams, then scheduling these videos to worker devices. The devices perform video analytics with these videos and then return the results to the master. Although this figure depicts a network of mobile devices processing video files, it is feasible for a single device to perform near real-time video analytics on dash cam video with the use of optimisations.

The specific contributions of this work are:
\begin{itemize}
    \item We have devised four optimisation techniques to overcome resource heterogeneity and constraints.
    \item We have implemented EDA on Android effectively interweaving several pre-mature and incompatible tools and libraries.
    \item We have demonstrated the feasibility of edge-based near real-time video analytics with a case study for two video analytics tasks: road hazard detection and driver distractedness detection.
    \item We have evaluated EDA with four mobile devices in five different network configurations.
\end{itemize}

\chapter{Background and Literature Review}\label{chap:background}
\section{Background}\label{sec:background}
EDA is based on technologies that have made various innovations in recent years. Such technologies include cloud computing, the internet of things, and edge computing.

\subsection{Cloud Computing}\label{sec:cloud_computing}
Cloud computing is a cheap and flexible method of using computer resources on demand. Without cloud computing, businesses that need computer resources would have to purchase and maintain all of the hardware by themselves. Even short-term needs for computer resources would require significant investment in hardware. Cloud computing allows businesses to pay for access to servers for a specified period of time. It is easy to scale, a business simply needs to pay for more resources if they are required to scale up. When more resources are provisioned than what is needed, costs can be reduced by deprovisioning excess resources in order to scale down.

Cloud computing providers offer different service models depending on the degree of abstraction or customisation required. Common service models include: infrastructure as a service (IaaS), platform as a service (PaaS), and software as a service (SaaS). IaaS is the least abstracted model, where clients have arbitrary control over the software installed on a cloud server, including the operating system. The client can choose to install any kind of software and configure it according to their needs. With PaaS, cloud providers supply tools or programming languages that clients can use to create their own applications that are hosted by the cloud provider. SaaS is the most abstracted model, where cloud-hosted applications are made for the direct use of clients. SaaS applications are typically accessed through web clients and includes services such as email, text editors, and accounting software.

Cloud computing is certainly convenient, but it does come with downsides.
The most immediate shortcoming is the lack of control over hired resources. When someone outright purchases computer resources, they have total control over everything. They decide on all the components of a system, such as specific hardware, operating systems, and various software configurations. On the other hand, clients of cloud computing are restricted in their control, only being able to select from a range of options. However, lower-level service models such as IaaS do provide freedom in the use of software and their configuration.
There are also concerns over security and privacy. Since clients do not own the cloud resources that they hire, they are limited in the extent that they can protect their data. Cloud providers have physical control over hired resources, so they determine what is the minimum level of security.
Finally, there is the issue of latency. There may be a great distance between a client and the cloud servers they have hired. Regardless of whatever optimisations are applied, there will always be some amount of delay in transferring data across this distance. Such transfer delays may be inconsistent due to factors affecting internet routing outside of the control of cloud providers and clients. For latency-sensitive tasks, cloud computing is not an ideal choice.

\subsection{Internet of Things}\label{sec:internet_of_things}
Internet of things (IoT) refer to devices that support network connectivity. Examples of IoT devices include smart fridges, security cameras, and dash cams. IoT devices are typically embedded with sensors that are constantly collecting data. However, most IoT devices lack the computational resources that are needed to process this data. In order to gain insight through analysis of data produced by IoT devices, this data must be offloaded to devices with greater computational resources. IoT data could be offloaded to cloud servers, though they may also be offloaded to nearby edge devices.

An enormous amount of video data is constantly produced by dash cams across the world. Dash cams have gained international popularity in recent years~\cite{mehrishambujEgocentricAnalysisDashcam2019,parkMotivesConcernsDashcam2016}, with 1 in 5 Australians having them installed in their cars~\cite{allianzRiseDashCam2019}. One method of creating further value from these videos is through video analytics. Video analytics is a computationally intensive task due to the nature of video data. Video data is dense as it is comprised of many frames, sequences of still images. A 10 second video with 30 frames per second (FPS) has 300 frames. In order to perform exact computation of video analytics on this video, all 300 frames must be analysed for the desired task. Video data is often produced by sensor nodes such as dash cams that lack the resources to perform video analytics. Thus, sensor nodes must transmit their video data to devices with greater resources in order to perform video analytics.

\subsection{Edge Computing}\label{sec:edge_computing}
Edge computing is a type of distributed computing where data is processed physically close to its source. Many data sources, such as IoT devices, lack the computational resources to process the data they produce. These devices must offload their data to other devices that have the computational capacity needed to process said data. This data could be offloaded to cloud servers, edge devices, or any number of other machines. Edge computing can supplement other kinds of distributed computing such as cloud computing, or it can completely replace it, depending on the circumstance. Most advantages of edge computing over cloud computing are due to the short distance between data sources and edge devices, in contrast to the potentially vast distances between data sources and cloud servers. This can result in edge computing having much less latency and bandwidth consumption than cloud computing. Although edge devices have more computational resources than the data sources they are associated with, they typically have much less computational resources than what is available with cloud servers. Therefore, edge computing is most suitable for tasks that require low latency and bandwidth usage, yet which do not have extremely high computational resource requirements.

Edge computing is a natural choice for video analytics due to the large size of video data. Transmitting all video data from a source to cloud servers would result in high bandwidth consumption and latency. Edge computing mitigates these problems by processing the video data close to the source. Fortunately, many modern dash cams support Wi-Fi connectivity so that their videos may be accessed by smartphone apps. This can allow a smartphone to act as an edge device by downloading and analysing the dash cam's videos. However, while modern smartphones generally have a greater processing capability than dash cams, they are still greatly limited when compared to non-mobile computers. A number of works have attempted to make up for this deficit by distributing tasks among a local network of smartphones~\cite{besencziFrameworkDistributedProcessing2014,dangeSchedulingTaskCollaborative2016,fernandoMobileCrowdComputing2012,habakkarimFemtoCloudsLeveraging2015}.

\section{Mobile Device Processing}\label{sec:mobile_device_processing}
In recent years, researchers have performed various machine-learning based tasks on mobile devices. This has been enabled by increasing interest in the use of machine-learning and the development of mobile-focused libraries such as TensorFlow Lite~\cite{googleTensorFlowLiteML2021}.

A FemtoCloud~\cite{habakkarimFemtoCloudsLeveraging2015} is a local network where a control device coordinates and schedules tasks to a group of mobile phone workers. It has been evaluated for processing tasks such as object recognition and running video games. While FemtoCloud and EDA share many similarities, the key difference is that FemtoCloud requires the use of a dedicated controller, whereas any EDA node may act as a master.

Machine learning systems using smartphone sensors have been used for medical purposes such as detecting Parkinson's~\cite{stamateDeepLearningParkinson2017}, assisting the visually impaired by alerting them to obstructions~\cite{delahozRealtimeSmartphonebasedFloor2017}, and assessing retinal disease~\cite{besencziFrameworkDistributedProcessing2014}. The last instance~\cite{besencziFrameworkDistributedProcessing2014} utilises an offline smartphone network framework with a scheduling algorithm that takes hardware information into account, such as available RAM and battery levels. The main similarity between this framework and EDA is that one node is responsible for gathering data.

Some virtual and augmented reality applications on smartphones offload their processing to edge servers~\cite{ranDeepDecisionMobileDeep2018,liMUVRSupportingMultiUser2018}. However, one augmented reality facial recognition application captures video data from a Google Glass, which is then offloaded to a smartphone for processing~\cite{mandalWearableFaceRecognition2015}.

\textcite{koukoumidisSignalGuruLeveragingMobile2011} developed an application whereby a network of dash-mounted smartphones can predict traffic light changes. The system uses a lightweight algorithm to identify traffic light signals from the smartphone's camera feed. Smartphones in nearby vehicles share their detection results with one another. The accumulated detection results are then passed to a support vector regression model that will predict the time at which upcoming traffic lights will change. The main similarity between EDA and this system is the use of a distributed network of mobile devices to process vehicle-based video data. Since this system only transmits detection results, it is able to utilise a network that covers a much greater distance than EDA, which transmits whole video files to nearby devices.

\section{Vehicular Video Analytics}\label{sec:vehicular_video_analytics}
The use of video analytics in conjunction with vehicles can serve a number of roles, the majority of which are safety-focused. However, this area of research comes with challenges, mostly due to the mobile nature and restricted space of vehicles. As dash cams lack the resources to process their own videos and it is inconvenient or outright impossible to install a server in a vehicle, live video data must be processed by alternative means. Cloud servers are unfit for vehicle-based video analytics due to their high latency, while edge-based methods such as those used by EDA are more suitable thanks to their low latency.

Vehicular video analytics can be grouped into three different categories. The first category involves outward-facing cameras that are used to detect road hazards such as other vehicles, pedestrians, and general road obstructions. The second category uses inward-facing cameras that monitor the driver and detect distracted driver behaviour. The third categories combines inward and outward-facing cameras to perform video analytics tasks that involve the outside of a vehicle as well as its driver.

\subsection{Road Hazard Detection}\label{sec:road_hazard_detection}
The vast majority of vehicular object detection systems were designed to detect pedestrians~\cite{ahmedPedestrianCyclistDetection2019,brunettiComputerVisionDeep2018,hurneyReviewPedestrianDetection2015} or other vehicles~\cite{mukhtarVehicleDetectionTechniques2015,sivaramanLookingVehiclesRoad2013,sunOnroadVehicleDetection2006}, while a much smaller number identify general road obstructions~\cite{creusotRealtimeSmallObstacle2015,moralesrosalesOnroadObstacleDetection2018,jiaRealtimeObstacleDetection2015}.

\subsubsection{Vehicle Detection}\label{sec:vehicle_detection}
Other vehicles are one of the common obstacles a driver will encounter. It is intuitive that a lot of research has been conducted on identifying vehicles.

\textcite{chadwickDistantVehicleDetection2019} created a ResNet-based model that is particularly suited to detecting distant vehicles.

\textcite{rybskiVisualClassificationCoarse2010} designed a system for identifying vehicles and their orientation. This was achieved by extracting features from video frames with a histogram of oriented gradients (HOG). The extracted features are then passed on to a support vector machine (SVM) to detect vehicles and their orientation. Unlike this system, EDA does not detect a vehicle's orientation. This feature was not found to be necessary to achieve EDA's task related to hazard detection.

\textcite{tehraniniknejadOnRoadMultivehicleTracking2012} also used a HOG and SVM to detect vehicles, though their system is capable of tracking multiple vehicles.

\textcite{zhangRealtimeVehicleDetection2018} utilised a HOG alongside the AdaBoost classification model and Kalman filter model to detect vehicles. Again, a HOG extracted features from video frames. These features are passed on to an AdaBoost model to identify vehicles, while the Kalman filter enables the tracking of said vehicles across adjacent frames. The authors claim that their system is capable of producing real-time results at a frame rate of 16 FPS. However, they do not specify the hardware that they used to evaluate their system. EDA currently lacks the ability to track vehicles across adjacent frames, though this is a planned addition as mentioned in Chapter~\ref{chap:future_work}.

\subsubsection{Pedestrian Detection}\label{sec:pedestrian_detection}
Pedestrians are another common obstacle, however, the detection of pedestrians differs from the detection of vehicles. 

\textcite{dominguez-sanchezPedestrianMovementDirection2017} created a framework for detecting pedestrians and their direction of movement with convolutional neural networks (CNN). They evaluated their framework with three different CNNs: AlexNet, GoogLeNet, and ResNet-10. With the fastest model, ResNet-10, they achieved a processing speed of 18 FPS when executed on a NVIDIA GTX 1070 GPU.

\textcite{tomeDeepConvolutionalNeural2016} designed a pedestrian detection system utilising CNN models. The system was evaluated on a desktop PC with a NVIDIA GTX 980 GPU. While their first CNN model was slow with a processing speed under 2 FPS, their optimised CNN model was much faster with a processing speed of 21.7 FPS. Similar to EDA, this system required various optimisations in order to reach near real-time processing speeds. However, while this system utilises a powerful GPU, EDA is restricted to the relatively weaker hardware of mobile devices.

\textcite{liuRobustFastPedestrian2013} developed a method of pedestrian detection for use at night. This method involves the use of SVM classifiers and feature extraction through pyramid entropy weighted HOGs. When evaluated on a desktop PC, the method ran at an average of 5 FPS.

\subsubsection{General Obstruction Detection}\label{sec:general_obstruction_detection}
\begin{figure}[t]
    \centering
    \includegraphics[width=0.72\textwidth]{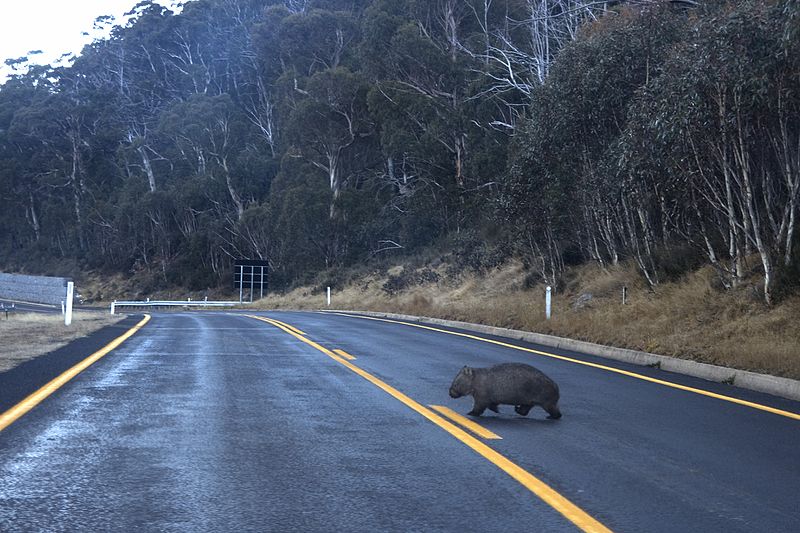}
    \caption{A wombat crossing a road, a potential hazard to drivers~\protect\cite{kozlenkoWombatCrossingRoad2012}.}
    \label{fig:wombat_crossing}
\end{figure}

The detection of general obstructions, such as a wombat crossing a road as depicted in Figure~\ref{fig:wombat_crossing}, differs significantly from the detection of vehicles or pedestrians. Instead of focusing on a consistent shape, many different shapes must be taken into consideration.

\textcite{creusotRealtimeSmallObstacle2015} designed a system for real-time detection of small obstructions on the road in front of the host vehicle. This system uses a restricted Boltzmann machine neural network to reconstruct the road's appearance, which helps to identify small anomalous objects. The test dataset comprised of videos recorded by the authors as well as videos downloaded from YouTube. A desktop PC was utilised to evaluate this system with videos of various resolutions. Additionally, some tests were performed with a mask applied to video frames, so that only the portion of frames that depict the road were processed. The fastest processing time of 103ms was achieved with masked videos at a resolution of 1360$\times$768, resulting in a frame rate just below 10 FPS.

\textcite{moralesrosalesOnroadObstacleDetection2018} created a general on-road obstacle detection system that is intended to work well in real-world conditions. An extended Kalman filter is used to detect and track objects in each frame. The test dataset consisted of videos recorded by the authors with a dash-mounted webcam. The system was evaluated on a low resource desktop PC, with a 2GHz CPU and 2GB of RAM. This system was found to be fairly slow, as it took an average of 960ms to process a single frame. Without significant optimisations, this system would be unable to operate in a near real-time manner as EDA is able to.

\textcite{jiaRealtimeObstacleDetection2015} developed a method of detecting on-road obstacles by comparing adjacent frames. Named as the ``two consecutive frames model'', this method uses the motion features present in consecutive frames to differentiate obstacles from the road itself. Evaluation was conducted on a laptop PC with videos recorded by the authors in addition to two other video datasets, KITTI~\cite{geigerAreWeReady2012} and CamVid~\cite{brostowSemanticObjectClasses2009}. This method was able to process frames in under 50ms, roughly equivalent to 15 FPS.

\subsection{Driver Distractedness Detection}\label{sec:driver_distractedness_detection}
When drivers perform actions that divert their focus away from driving, they greatly increase the likelihood of causing an accident~\cite{dingusDriverCrashRisk2016}. It would be helpful to curb this behaviour by automating its detection through automated processes. Such automated detection has been achieved through video analytics, where machine-learning techniques are applied to video data from inward-facing cameras. Some of these systems are based on estimating the driver's gaze to determine if they are paying attention to the road~\cite{leeRealTimeGazeEstimator2011,choiRealtimeCategorizationDriver2016,voraDriverGazeZone2018}, while others detect actions unrelated to driving such as talking on the phone~\cite{yanDriverBehaviorRecognition2016,tranRealtimeDetectionDistracted2018,kapoorRealTimeDriverDistraction2020,huangHCFHybridCNN2020}.

\textcite{yanDriverBehaviorRecognition2016} developed a driver behaviour recognition system that identifies if the driver performs actions such as talking on the phone, eating, or smoking. The system extracts skin-like regions from video frames with a Gaussian mixture model which are then passed on to a R*CNN model for classification. The system was evaluated on a desktop PC, however, the authors did not provide results on processing time.

\textcite{tranRealtimeDetectionDistracted2018} created a framework that alerts drivers when it identifies distracted driving behaviour. The framework passes each frame of a video to a CNN model that determines what kind of action the driver is performing. Several CNN models are compatible with the framework. The authors demonstrated this compatibility by evaluating it with VGG-16, AlexNet, GoogLeNet, and ResNet. To evaluate the framework, the authors constructed a testbed that simulates a driving environment while recording the driver. In addition to the cameras, the testbed included a NanoPi M3 running Android that delivered alerts to the driver, and a NVIDIA Jetson TX1 developer kit which processed video frames. The VGG-16 model was found to have the fastest processing speed of 14 FPS. While a mobile Android device was involved in this project, all video processing was performed on a powerful developer kit. On the other hand, EDA achieved a similar task with all processing being performed by mobile devices.

\textcite{kapoorRealTimeDriverDistraction2020} designed a driver distraction detection system that identifies actions such as eating or texting. This system was implemented on a software defined cockpit powered by Android in-vehicle-experience, though the authors omitted specific hardware details. The authors compared the results of running their system with several CNN models. These models were MobileNetV1, MobileNetV2, InceptionV3, and VGG-16. MobileNetV1 was found to be the most efficient, so it was further optimised to process video data faster and with greater accuracy. The authors claim that their system can process in real-time with this fine-tuned model, however, they do no provide processing time results.

\textcite{huangHCFHybridCNN2020} developed a hybrid CNN framework for distracted driver recognition. This framework combines the ResNet-50, InceptionV3, and Xception models. The models are first used to extract behaviour features from video frames, then converts them one-dimensional vectors, finally classifying the detected behaviour. Evaluated on a desktop PC, the framework was able to process frames in 41ms on average, roughly equivalent to 24 FPS.

\clearpage
\subsection{Simultaneous Inner/Outer Analysis}\label{sec:simultaneous_inner_outer_analysis}
A few papers have utilised the analysis of both inner and outer vehicular video feeds. While they bear a resemblance to EDA, none have been found that utilise a distributed network of mobile devices in a similar manner to EDA.

\textcite{trivediLookingInLookingOutVehicle2007} discussed various potential video analytics tasks such as: estimating the driver's head-pose by processing the inner video with a hidden Markov model, applying a Kalman filter to the outer video to detect and track lanes, as well as using both inner and outer video to predict lane-changing.

\textcite{jainCarThatKnows2015} applied an auto-regressive input-output hidden Markov model on inner and outer video data as well as GPS data to predict predict future lane changes and turns. The authors state that this work could be incorporated within an advanced driver assistance system that could warn drivers of potentially dangerous manoeuvres. This system utilises multiple types of sensor data in addition to video data, whereas EDA only processes video data.

\textcite{rezaeiLookDriverLook2014} designed a system whereby inner and outer video data is analysed to predict the risk of an accident. Asymmetric appearance modelling is applied to inner video to detect drowsiness or distractedness, while global Haar classifiers are used with outer video to identify vehicles and their distance from the host vehicle. The outputs of these processes are combined through fuzzy fusion to determine the level of risk, alerting the driver when risk meets a certain threshold.

\chapter{EdgeDashAnalytics}\label{chap:eda}
In this chapter, we present EdgeDashAnalytics (EDA) in detail. In particular, we describe the design and implementation of EDA in the context of an Android mobile app.

\begin{figure}[htb]
    \centering
    \includegraphics[width=0.9\textwidth]{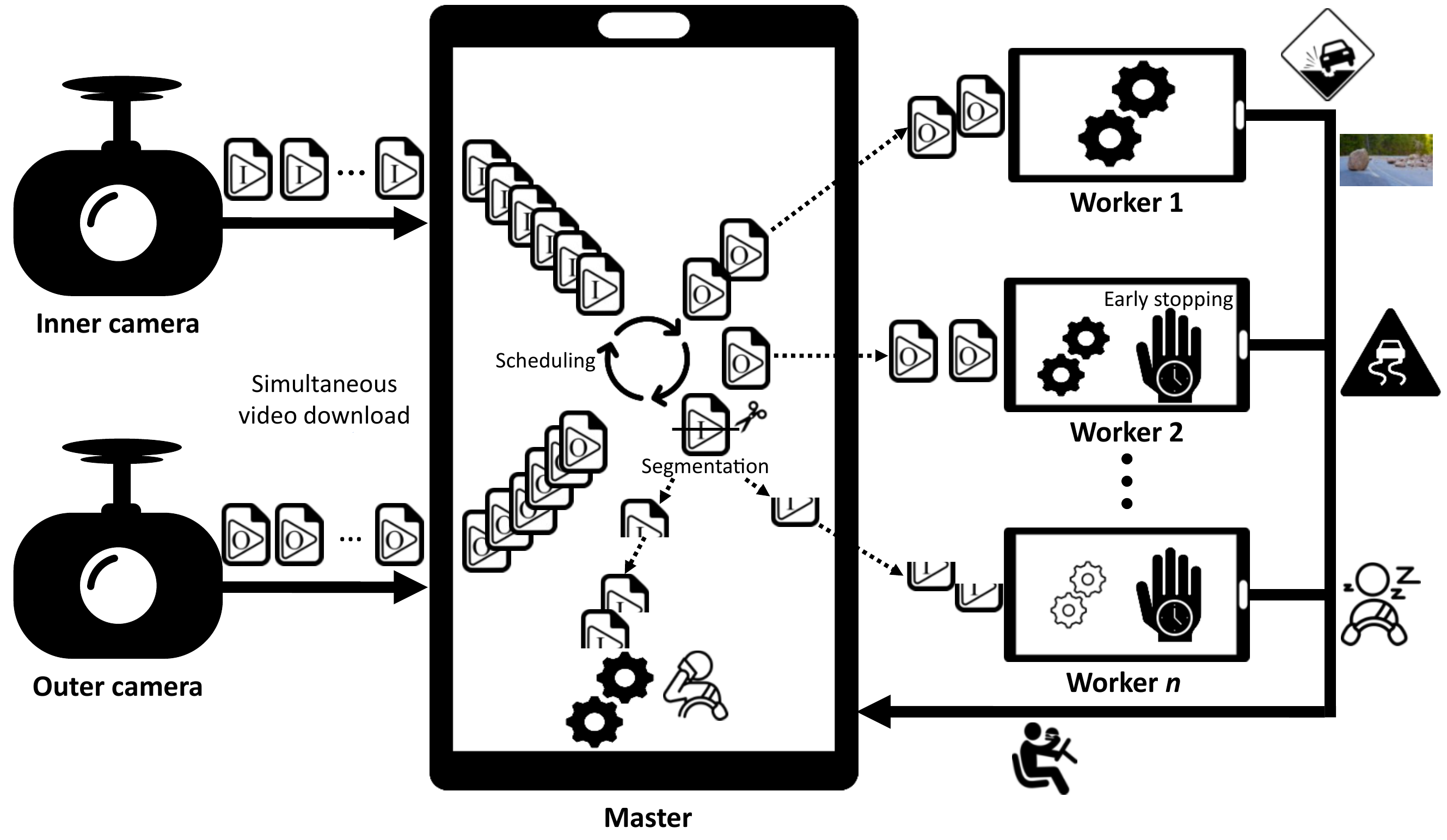}
    \caption{Detailed overview of EDA.}
    \label{fig:overview_detailed}
\end{figure}

\clearpage
\section{Overview}\label{sec:overview}
EDA consists of four key functionalities: simultaneous video download and analysis, scheduling, segmentation and early stopping (Figure~\ref{fig:overview_detailed}).
Currently, EDA is designed to deal with two dash cams simultaneously. In particular, one captures the outer forward view of the vehicle, while an inner camera records the driver. Fortunately, many commercial dash cams are packaged with a secondary camera, avoiding the need to purchase an entirely separate dash cam.
Therefore, a master mobile device downloads the videos produced by each dash cam concurrently. The outer videos are analysed to identify road hazards, while instances of distracted driver behaviour are detected with inner videos. The analysis of these videos are finished before the next pair of videos are produced, thereby achieving near real-time turnaround.
Simultaneously downloading and analysing two separate sources of video data is computationally intensive. It would be difficult for a single mobile device to process this video data in a real-time manner. However, it is much more feasible to achieve near real-time processing by utilising the shared resources of a local network of mobile devices.

Whenever a video is downloaded by the master device, it makes a scheduling decision based on the video source and the circumstances of the network. Outer videos are prioritised for assignment to devices with the greatest processing capacity. This prioritisation is made as road hazards are likely to be more dangerous than distracted driver behaviour. Inner videos are assigned to the computationally weaker devices in the network. When there are at least two devices in the network, each video from the downloaded pair is assigned to a different device so that the videos are concurrently processed. Similarly, if there are at least three devices in the network, then videos are segmented to ensure that all devices are analysing videos simultaneously. This concurrent analysis helps to achieve the low turnaround of results necessary for near real-time processing.

In situations where EDA's network is comprised of too few devices, or the devices are computationally weak, videos may not be analysed fast enough to achieve near real-time processing. In such circumstances, the optimisation technique of early stopping ensures that videos are analysed at a speed that make it possible to reach near real-time processing. Essentially, early stopping works by associating the time taken to analyse a video with a period of time that is proportional to the video's length. If video analysis is not completed when this time is reached, then analysis is terminated early and the remainder of the unprocessed video is discarded.

\section{Design and Implementation}\label{sec:implementation}
In this section, we detail the design and implementation of EDA. The four key functionalities of EDA are actually implemented through the user interface, video analysis, networking, and the scheduling algorithm.

The Android operating system was chosen as the platform on which to develop EDA. This decision was made due to the widespread use of Android for various mobile devices, the abundance of supporting libraries, as well as familiarity with the Java programming language. However, Android applications have many constraints on how they may be implemented. Despite such constrains restricting the options available in developing EDA, it was still able to function according to its design.

\clearpage
\subsection{Data Objects}\label{sec:data_objects}
EDA utilises several objects to represent various types of data. These objects will be referred to throughout this chapter, and are described in the following list.

\begin{itemize}
    \item Video: represents a video file, stores the video's name and file path.
    \item Result: represents a result file, storing its file name and path.
    \item Command: an enum, instances are sent to devices to provide instructions. The commands are:
    \begin{itemize}
        \item ANALYSE: requests that the video paired with this command should be analysed.
        \item SEGMENT: similar to ANALYSE, requests that the paired video should be analysed and that it is a video segment.
        \item COMPLETE: indicates that the file transfer of an associated video was successful.
        \item RETURN: paired with a result file.
        \item HW\_INFO\_REQUEST: a command requesting that the receiver responds with hardware information.
        \item HW\_INFO: indicates that the message contains JSON-encoded hardware information.
    \end{itemize}
    \item Message: container class holding a Command instance and a Video or Result instance.
    \item Endpoint: the Nearby Connections API identifies devices through strings known as endpoint IDs. Endpoint objects store this ID, along with a unique human-readable name and a boolean value indicating connection status.
    \item HardwareInfo: class representing the hardware information of a device. It contains a device's: CPU frequency, number of CPU cores, total RAM, currently available RAM, total storage, currently available storage, and current battery level.
    \item Payload: a class defined by the Nearby Connections API that represents messages. They are identified through long payload IDs. Two types of payload are used by EDA, file and byte.
    \begin{itemize}
        \item File payload: contains a file that is intended for delivery using the Nearby Connections API. File payloads are created using files, and a file can be extracted from a file payload.
        \item Byte payload: contains an array of bytes. These are used to send colon-delimited strings which are composed of commands, payload IDs, and filenames (i.e. ``command:ID:filename'').
        \begin{itemize}
            \item The command is a string representation of a Command object. This indicates what should be done with a paired file payload, or a standalone instruction if sent alone.
            \item The payload ID is copied from a file payload that is paired with this byte payload. This ID is used as the key for several dictionaries used throughout the transfer process to match the file payload and byte payload together. This field is omitted for command messages that are sent alone.
            \item The filename is the original name of the video file. This must be sent in a byte payload as file payloads do not preserve filenames.
        \end{itemize}
    \end{itemize}
\end{itemize}

\begin{figure}
    \centering
    \begin{subfigure}{0.26\textwidth}
        \centering
        \includegraphics[width=\textwidth]{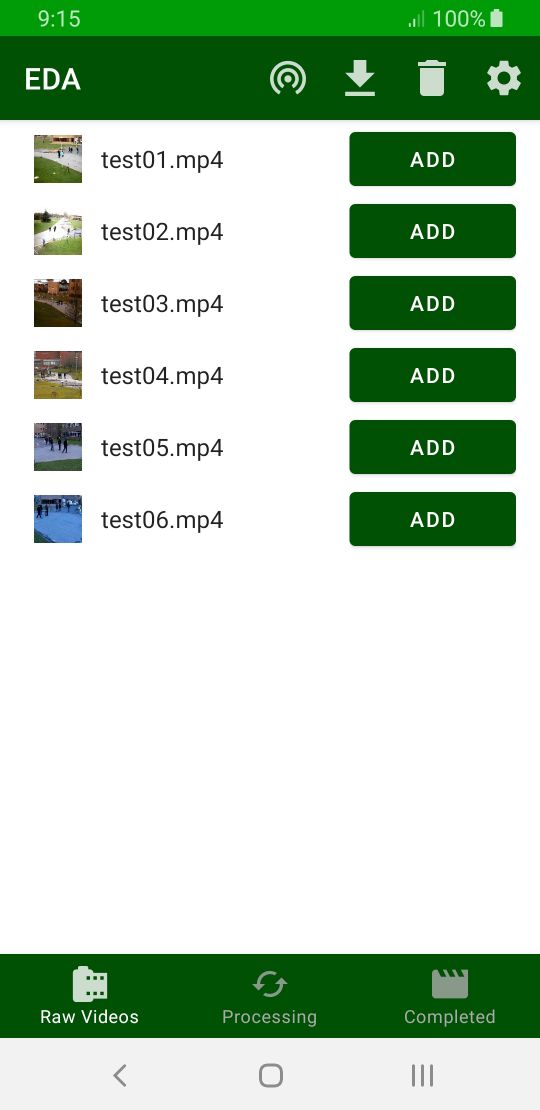}
        \caption{Raw video list.}
        \label{fig:videos_raw}
    \end{subfigure}
    \hfill
    \begin{subfigure}{0.26\textwidth}
        \centering
        \includegraphics[width=\textwidth]{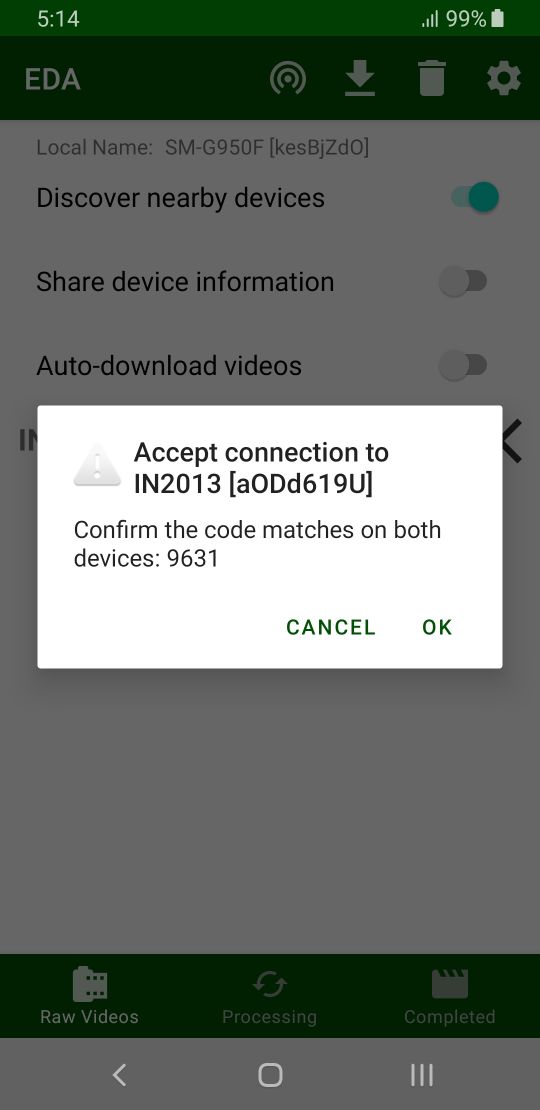}
        \caption{Connection prompt.}
        \label{fig:connection_prompt}
    \end{subfigure}
    \hfill
    \begin{subfigure}{0.26\textwidth}
        \centering
        \includegraphics[width=\textwidth]{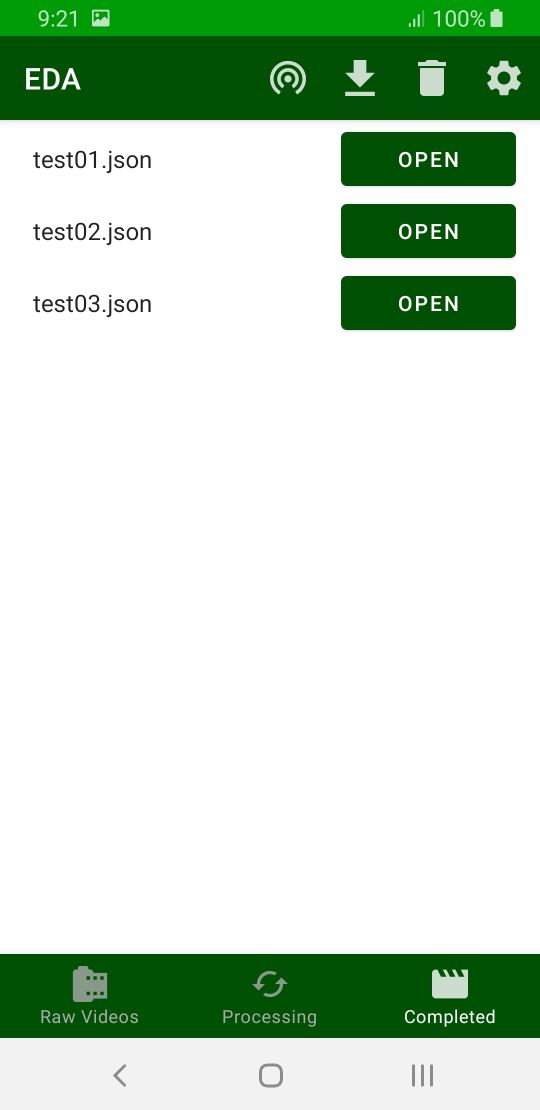}
        \caption{Results list.}
        \label{fig:results}
    \end{subfigure}
    \caption{Screenshots of EDA in operation.}
    \label{fig:screenshots}
\end{figure}

\subsection{User Interface}\label{sec:user_interface}
EDA's user interface informs the user which videos are in the network and their current state. This is managed by three lists: raw videos, processing videos, and results. The raw list shows all of the unprocessed videos that are stored on the device. Clicking on a video in the raw list will manually initiate scheduling with this video. The processing list displays all videos that are currently being processed, meaning that they are being transmitted or analysed. The results list includes all of the results files produced from video analysis. Clicking on an entry in the results list will open and display the results file. Additionally, connecting to other devices is managed through a menu in the UI.

\subsubsection{Navigation and Preferences}\label{sec:navigation_and_preferences}
The UI is set up by MainActivity which creates fragments and handles fragment selection. Interactable UI elements include a BottomNavigationView used to select list fragments and a Menu used to open settings and the ConnectionFragment. The BottomNavigationView contains three clickable icons, these are used to select the raw videos, processing videos, and results fragments. The connection button also displays a fragment used for connectivity actions, it is located in the Menu instead of the BottomNavigationView as it serves a different purpose than the list fragments. User preferences are set via the interactable elements in SettingActivity which is also displayed through a button in the Menu. The behaviour of preference widgets are defined in Android's preference library, they are simply created in the root\_preferences layout file. The values set by these preference widgets are accessed through the a SharedPreferences object whenever they are needed.

\subsubsection{List Fragments}\label{sec:list_fragments}
VideoFragment implements the UI for displaying and interacting with videos, while ResultsFragment does the same for result files. There are two VideoFragment instances, one for raw videos and one for processing videos. As seen in Figure~\ref{fig:videos_raw}, these VideoFragments are displayed as a list where each item includes a video's filename, thumbnail, and a button to manually initiate analysis. MainActivity uses a BottomNavigationView to switch between these fragments. The BottomNavigationView contains three buttons, one for each fragment. Pressing a button will hide the current fragment and will display the selected fragment through calls to a FragmentManager. The layout of these list fragments consist of a RecyclerView list which is managed through an adapter subclass, either VideoRecyclerViewAdapter or ResultRecyclerViewAdapter.

Video files are represented through Video objects which stores values such as the video's file path and MediaStore ID. Each VideoFragment contains a VideoRepository which manages a MutableLiveData instance. Video objects are stored in these MutableLiveData instances, and they are accessed and updated through other classes such as VideoRecyclerViewAdapter. A video is added or removed from a particular VideoFragment by posting AddEvents or RemoveEvents to the EventBus along with the Video in question and an enum identifying which VideoFragment the event is intended for. These events are received by a VideoEventHandler that updates a MutableLiveData instance and the UI according to the contents of the message. For example, posting a Video with an AddEvent and an enum type of RAW will make the video appear in the rawFootageFragment. ResultsFragment is structured and behaves similarly to VideoFragment, except Result objects take the place of Video objects.

The networking and video management processes must be active at the same time for them to operate together. Since only a single Activity can be active at a time, it was decided to extend Fragment instead, since Android allows multiple active fragments.

\subsection{Video Analysis}\label{sec:video_analysis}
This project demonstrates two video analytics tasks, one for outer videos and one for outer videos. Outer videos are recordings made by an outward-facing dash cam that captures the area in front of the host vehicle. Inner videos are created by an inward facing dash cam which captures the host vehicle's driver. When a video is processed, its frames are extracted and analysed according to the type of video. Upon completion of a video's analysis, the position and category of detected objects and their flags are written to a JSON results file.

Video analysis is handled by the VideoAnalysis class and its subclasses OuterAnalysis and InnerAnalysis. The VideoAnalysis class instantiates a MediaMetadataRetriever object from the standard Android library which it uses to extract video frames with its getFrameAtIndex method. Each frame is extracted as a Bitmap object which is first downscaled to match the input dimensions of the subclass's model, then is passed to subclass's processFrame method. This downscaling does decrease detection accuracy, but is necessary in order to reduce processing time to real-time speeds. As the subclass names suggest, OuterAnalysis is responsible for analysing outer videos, while InnerAnalysis is responsible for analysing inner videos. Both of the subclasses pass video frames to TensorFlow Lite libraries which performs image recognition on them. The results from the libraries are then analysed to determine whether or not to flag an object. These results are then written to a JSON file.

Unfortunately, none of the Android machine learning libraries appear to directly support processing video files, though some offer limited support for video streams. Many papers on mobile device video analytics appear to use custom frameworks built for that particular work's purposes. This creates the need for a lot of extra work that would not be necessary if there were a general framework for machine learning based video processing. As the focus of this project is on distributed processing rather than machine learning techniques, we decided to use relatively simple video analytics tasks.

\subsubsection{Outer Analysis}\label{sec:outer_analysis}
The outer video analysis task is to detect potential road hazards and identify if the driver is tailgating. This involves identifying objects on the road with an object detection model. The lower-middle area of a video is marked as the road. If any non-vehicle objects are detected in this area, then they are flagged as hazards. Additionally, any vehicles that are identified as being too close to the host vehicle are flagged as hazards due to potential tailgating.

OuterAnalysis uses the TensorFlow Lite task library~\cite{googleTensorFlowLiteTask} to detect potential hazards. It first instantiates an ObjectDetector object with the selected object detection model, such as the lightweight MobileNetV1~\cite{tensorflowSSDMobileNetV12021}. The Bitmap passed to processFrame is converted to a TensorImage which is passed to the ObjectDetector's detect method that returns a list of Detection objects, representing all of the objects detected within a frame. Non-vehicle objects that are detected on the road are flagged as potential hazards. Vehicle objects that are large enough to indicate they are very close to the host vehicle are flagged for potential tailgating. Once every frame within a video is processed, the results are written to a JSON file that is structured as follows. The outermost component is an array of frame objects, each of these objects contain the frame's index and an array of detected objects. These detected objects consist of the object's category, a boolean denoting if the object is a potential danger, the confidence score of the detection, and the object's bounding box, represented as four integers denoting the bounding box's bottom, left, right, and top edges.

\subsubsection{Inner Analysis}\label{sec:inner_analysis}
InnerAnalysis uses the TensorFlow Lite support library~\cite{thetensorflowauthorsTensorFlowLiteSupport2021} to detect driver distractedness. It first instantiates an Interpreter object with the selected pose estimation model, such as MoveNet Lightning~\cite{tensorflowMoveNetLightning2022}. The frame Bitmap accepted by processFrame is converted to a TensorImage which is passed to the Interpreter's run method along with an output TensorBuffer. This TensorBuffer is filled with a float array that identifies the body parts present within a frame as well as their coordinates. Distractedness is then determined by identifying if the driver's hands or eyes are not focused on driving. If a hand is above three-quarters of the frame height, such as when a driver holds their phone to their ear, then the driver is flagged as distracted in this frame. If the eyes are positioned downwards relative to the ears, such as a when a driver is glancing at their phone, then the driver is flagged as distracted in this frame. Once every frame within a video is processed, the results are written to a JSON file that is structured as follows. The outermost component is an array of frame objects, each of these objects contain the frame's index, a boolean denoting driver distractedness, and an array of body part objects. The objects consist of the body part category, the confidence score of its identification, as well as its X and Y coordinates.

\clearpage
\subsubsection{Analysis Results}\label{sec:analysis_results}

\begin{figure}[htb]
    \centering
    \begin{subfigure}{0.49\textwidth}
        \centering
        \includegraphics[width=\textwidth]{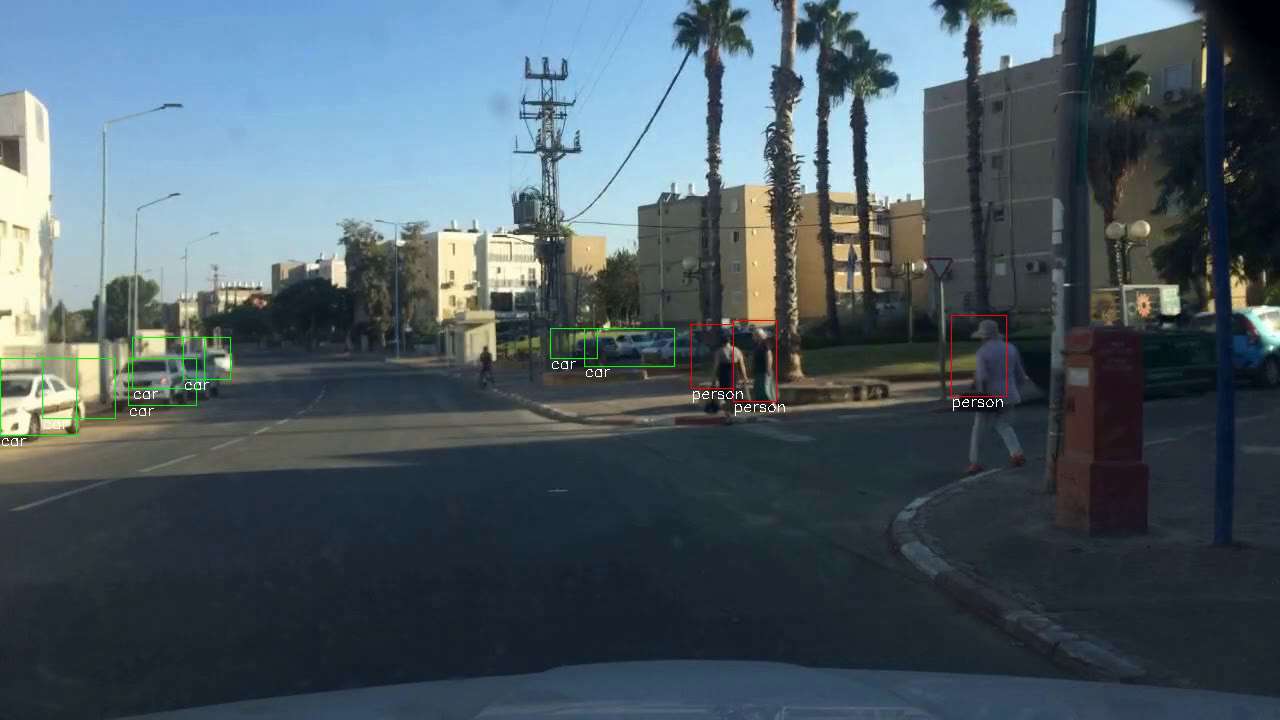}
        \caption{MobileNetV1 detection results.}
        \label{fig:mobilenet_boxes}
    \end{subfigure}
    \begin{subfigure}{0.49\textwidth}
        \centering
        \includegraphics[width=\textwidth]{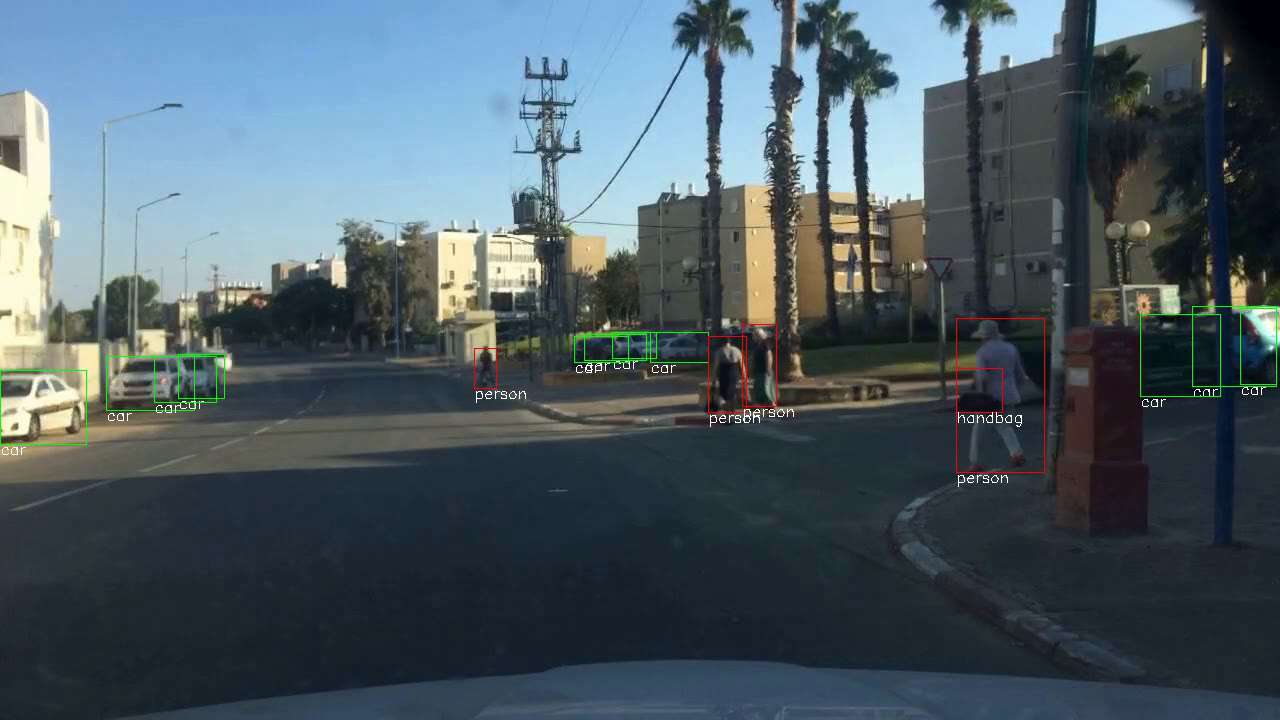}
        \caption{EfficientDet\hyp{}Lite4 detection results.}
        \label{fig:efficientdet_lite4_boxes}
    \end{subfigure}
    \caption{Comparison of outer detection results between MobileNetV1 and EfficientDet\hyp{}Lite4 models.}
    \label{fig:boxes}
\end{figure}

The result files can be visualised with the use of simple python scripts\footnote{Available in GitHub repository: \url{https://github.com/JaydenKing32/EdgeDashAnalytics}}. The outer visualisation script reads each frame of the original video file and draws bounding boxes at the coordinates specified in the result file. These bounding boxes surround detected objects and include category labels such as ``person'' or ``car''. The colour of the bounding boxes are determined by the hazard classification. Potential hazards have red bounding boxes, while non-hazards have green bounding boxes. An example of this is shown in Figure~\ref{fig:boxes}, which also illustrates the difference in results produced by different models. Figure~\ref{fig:mobilenet_boxes} shows the output from the lightweight MobileNetV1~\cite{tensorflowSSDMobileNetV12021} model, which contains several obvious errors. Its bounding boxes are incorrectly sized or are offset, it completely missed the cyclist, and misidentifies a single car as two separate cars. In comparison, the output produced by the larger EfficientDet\hyp{}Lite4~\cite{tensorflowEfficientDetLite42021} model shown in Figure~\ref{fig:efficientdet_lite4_boxes} contains fewer errors. Its bounding boxes for people are correctly sized and positioned, though it appears to have some difficulty distinguishing between multiple cars when they are close together. However, this increased accuracy comes at the cost of greater computational load, leading to EfficientDet\hyp{}Lite4 taking much longer to process videos than MobileNetV1. Despite its inaccuracy, MobileNetV1 is able to produce results in a real-time manner whereas EfficientDet\hyp{}Lite4 cannot. Therefore, the lightweight MobileNetV1 model was used for evaluation instead of larger models such as EfficientDet\hyp{}Lite4.

\begin{figure}[hbt]
    \centering
    \includegraphics[width=0.8\textwidth]{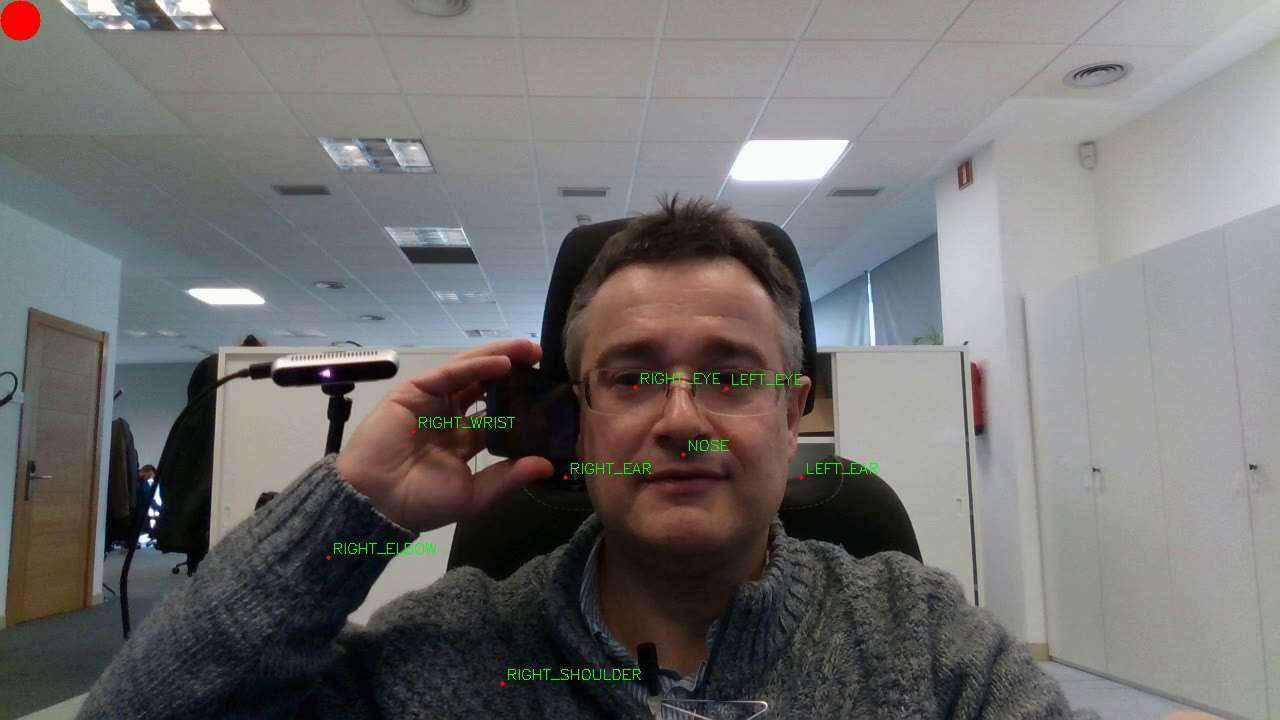}
    \caption{Example of visualised inner results.}
    \label{fig:inner_example}
\end{figure}

The visualisation of inner results is demonstrated by Figure~\ref{fig:inner_example}. A coloured dot is placed at the coordinates of each detected body part, with the body part's name written beside it. In addition, every frame contains a coloured circle in the top-left corner that signals the distractedness of the driver. Red indicates that the driver is distracted, while green denotes no distraction. The MoveNet Lightning~\cite{tensorflowMoveNetLightning2022} pose estimation model was used for this task for detecting the position of various body parts. It was designed to process images depicting the full body, instead of just the upper portion of the body that would be captured by an inward-facing dash cam. This results in reduced accuracy and issues such as incorrectly identifying off-screen body parts, as seen with the right elbow in Figure~\ref{fig:inner_example}. Despite such problems, the model serves its purpose well enough and is fast enough to produce results in a real-time manner.

\subsubsection{Early Stopping}\label{sec:early_stopping}
The method of early-stopping was created to ensure that videos are analysed in a time period shorter than the video's length, thereby achieving near real-time processing. This feature is controlled by the early-stop divisor (ESD) value. The ESD is used to divide a video's length, the result is the maximum running time for video analysis. If a video is still being analysed when this time is reached, then the analysis is stopped early and the remaining unanalysed portion of the video is discarded. Due to relatively lengthy operations such as file transfers, the effective ESD value must be greater than 1 in order to reach near real-time turnaround. However, it may not be necessary to set an ESD on devices with especially high processing capabilities, as they may be able to fully process videos in a near real-time manner without the use of early-stopping.

\subsection{Networking}\label{sec:networking}
The technology which was most often found when conducting research on smartphone edge networks was Wi-Fi Direct, so it seemed like a good decision to utilize a Wi-Fi Direct system for EDA. However, this proved to be difficult as documentation on Wi-Fi Direct for Android was sparse and outdated. Instead, the Nearby Connections API~\cite{googleNearbyConnections2021} was chosen for communication between smartphones as it had up-to-date documentation and was easier to implement. This API allows smartphones to transmit video files and command messages between one another via Bluetooth and Wi-Fi.

While the Nearby API sample code~\cite{googleOverviewNearbyConnections2018} worked fine on any Android device that had Google Play Services installed, they were all implemented into a single Activity. This was a problem as only one Activity can run at a time. A user could establish a P2P connection in one Activity, but switching to any other Activity would terminate said connection. This problem was solved through the use of Fragments which can run concurrently with one another. This allows a connection that is established within one fragment to persist through other fragments in the app.

\begin{figure}[htb]
    \centering
    \includegraphics[width=0.5\textwidth]{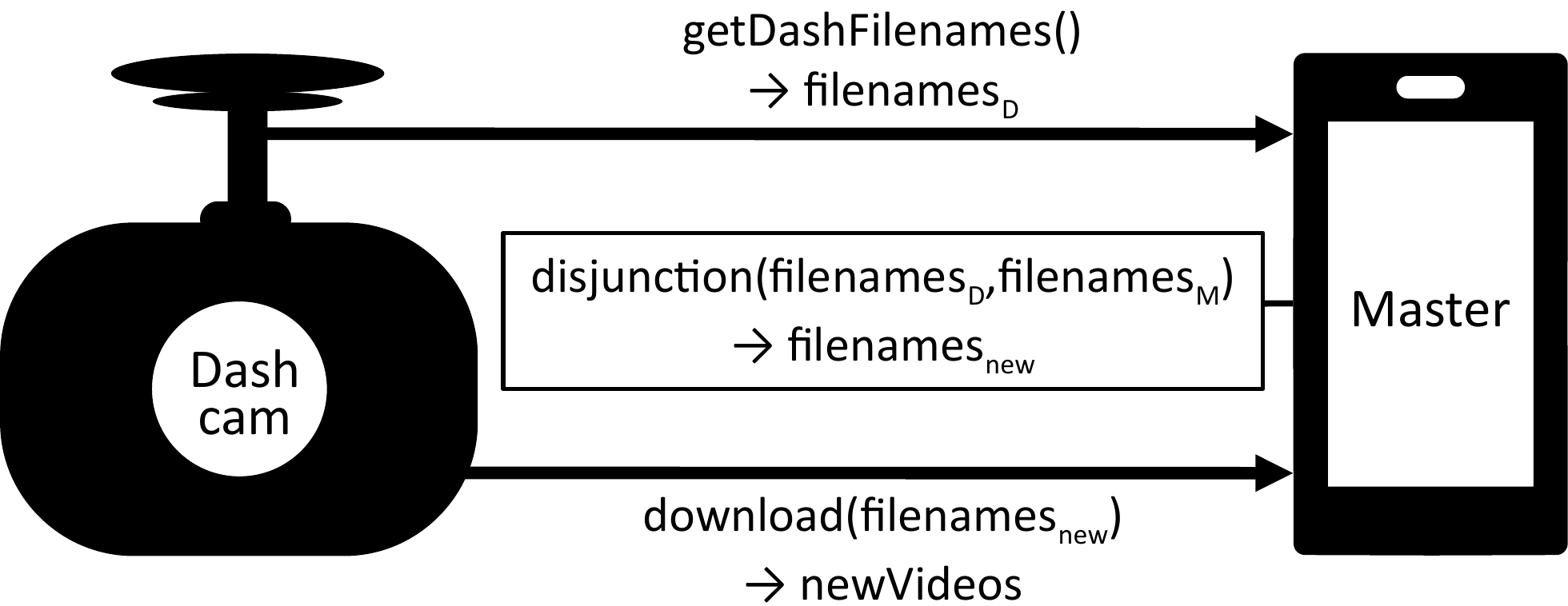}
    \caption{Downloading videos from dash cam.}
    \label{fig:dash_download}
\end{figure}

\subsubsection{Dash Cam Download}\label{sec:dash_cam_download}
Before videos can be downloaded, a device must connect to the dash cam via Wi-Fi. The dash cams act as a Wi-Fi access point, so mobile devices can connect to them just like a wireless router. This is currently achieved manually through the device's Wi-Fi settings, though this process could be automated in the future. As the tested dash cam does not offer a direct API, downloads must be processed through a HTML interface. The jsoup~\cite{hedleyJsoup2022} and Fetch~\cite{francisFetch2022} libraries are used for this purpose. The DashCam class passes the IP address of the connected dash cam to jsoup's connect method which returns a Document object representing the dash cam's HTML interface web page. This web page lists all of the videos recorded by the dash cam, so it is a simple matter of parsing the Document object to identify these videos and obtain their URLs. DashCam's downloadVideo method can then download these videos by passing their URLs to Fetch's enqueue method.

Upon initiation of automatic downloading, the startDashDownload method will pass the downloadLatestVideos method to a ScheduledExecutorService. downloadLatestVideos is illustrated in Figure~\ref{fig:dash_download}. It identifies which videos are new by performing a disjunction on the list of videos stored on the dash cam and the list recording the names of downloaded videos. downloadLatestVideos will then download these videos from the dash cam in sequence, starting with older videos. The videos are downloaded in concurrent pairs of outer and inner videos, waiting the duration of a video before initiating the next pair of downloads. On download completion, the device will make a decision based on the circumstances, as described in Section~\ref{sec:algorithm}.

The above describes using downloadLatestVideos to automatically download any arbitrary set of videos stored on the dash cam. However, for the purposes of consistent testing in evaluation, downloadLatestVideos is replaced with downloadTestVideos. downloadTestVideos works similarly to downloadAll, instead it downloads a predefined list of videos from the dash cam.

\subsubsection{Connection}\label{sec:connection}
The ConnectionFragment contains an advertising switch, a discovery switch and a device list. Advertising and discovery in the Nearby Connections API is structured through strategies defined by the library. The strategy used for advertising and discovery in EDA is P2P\_STAR, as it suits the app's star topology of one master and multiple worker devices. The device list is controlled through a deviceAdapter which stores a list of Endpoint objects, updates the UI, and handles user input.

\begin{figure}[htb]
    \centering
    \includegraphics[width=0.55\textwidth]{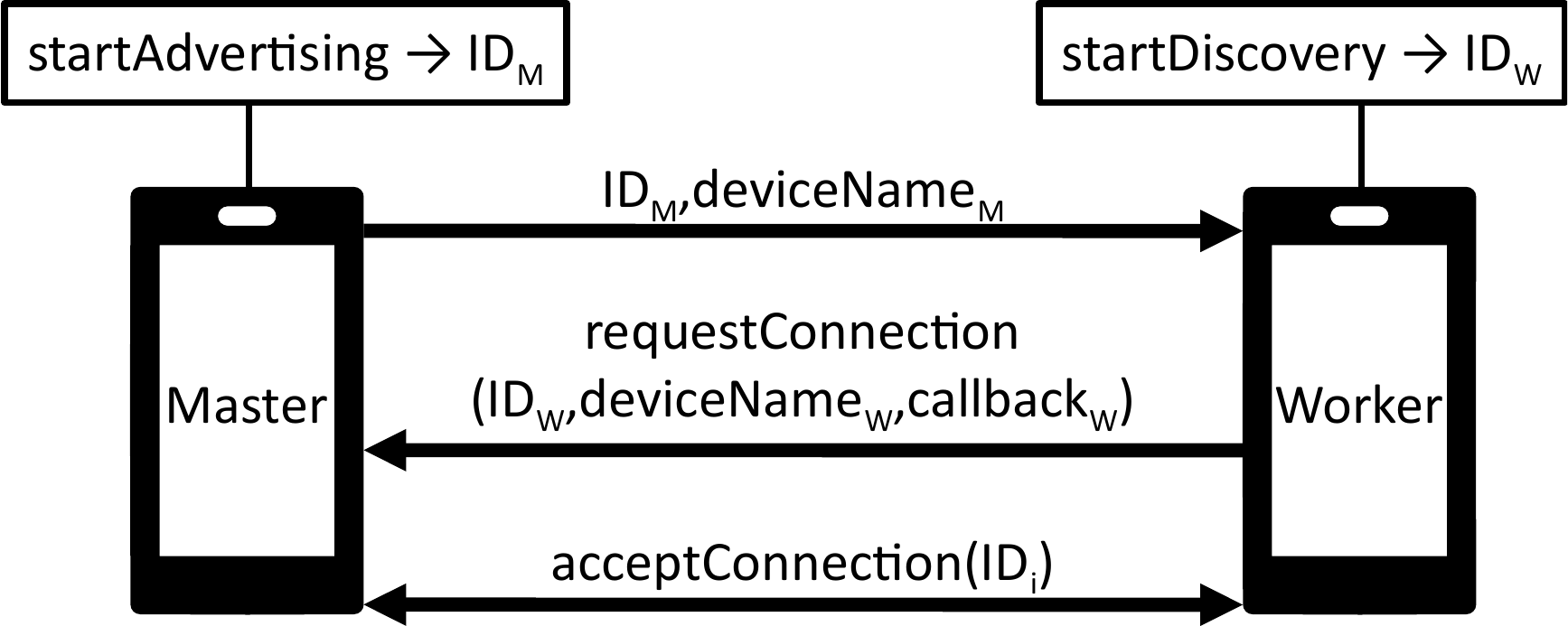}
    \caption{Connecting master and worker devices.}
    \label{fig:connection}
\end{figure}

Advertising is initiated by pressing the advertising switch when it is disabled. This calls the startAdvertising method of connectionsClient, passing P2P\_STAR and connectionLifecycleCallback, which makes the device start advertising itself. connectionLifecycleCallback responds to connection requests by displaying a message and a prompt to the user as shown in Figure~\ref{fig:connection_prompt}. If the user approves the connection then a PayloadCallback instance will be passed through acceptConnection call on connectionsClient, establishing a connection between the devices. Pressing the advertising switch when it is enabled simply calls the stopAdvertising method of connectionsClient, which makes the device stop advertising.

Pressing the discovery switch when it is disabled calls the startDiscovery method of connectionsClient, passing P2P\_STAR and endpointDiscoveryCallback, which makes the device start discovering. Upon endpoint discovery, endpointDiscoveryCallback creates an Endpoint object with the information provided by the endpoint and adds it to the discoveredEndpoints list. If an endpoint is lost, then the corresponding Endpoint object is removed from the discoveredEndpoints list. Adding or removing items from the discoveredEndpoints list updates the device list displayed in the ConnectionFragment by calling notifyItemInserted and notifyItemRemoved on the deviceAdapter. Pressing the discovery switch when it is enabled simply calls the stopDiscovery method of connectionsClient, which makes the device stop discovering.

Selecting a device in the ConnectionFragment's device list will initiate a connection with it by passing the selected device's endpoint to connectEndpoint. This process is illustrated in Figure~\ref{fig:connection}. The only thing that distinguishes master and worker devices is whether they advertise or discover. Master devices advertise themselves to worker devices, while worker devices discover master devices.

\subsubsection{File Transfers}\label{sec:file_transfer}
While the Nearby Connections API manages low-level networking operations, it does not offer a ready-made protocol for handling file transfers. A simple protocol was created for EDA, where transferred video and result files are bundled with a command message that instructs what should be done with the associated file.

\begin{figure}[htb]
    \centering
    \includegraphics[width=0.5\textwidth]{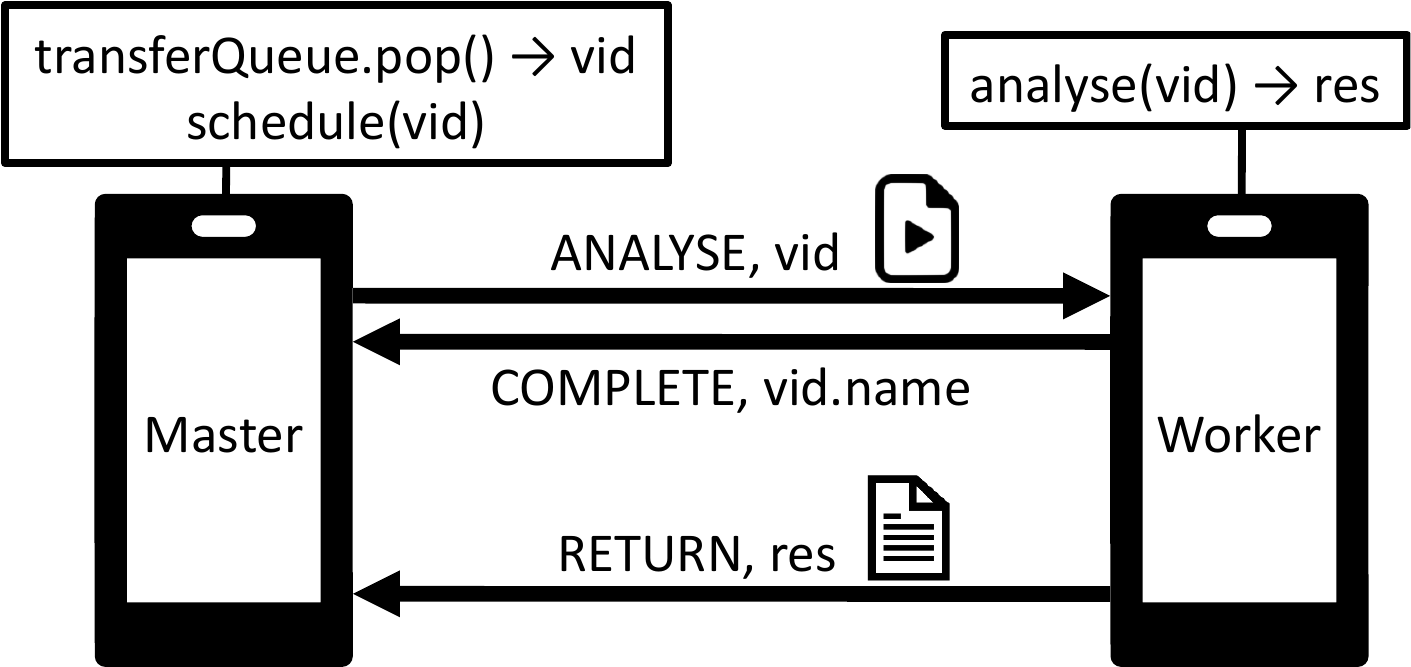}
    \caption{File transfer between master and worker devices}
    \label{fig:file_transfer}
\end{figure}

Video files can be transferred between devices manually or automatically. Manual transfers are initiated by clicking on videos listed in the rawFootageFragment. If EDA is not connected to any other devices, then selected videos will simply initiate local analysis. If EDA is connected to other devices, then Message objects containing the selected videos and ANALYSE commands are added to the transferQueue, then the nextTransfer method is called. If video segmentation is enabled, then the video will be split into equally sized segments, these segments are added to the transferQueue instead of the original video. The number of segments a video is split into can be set manually or automatically determined. nextTransfer removes the message from the head of the transferQueue and passes it to sendFile along with the endpoint ID of a device chosen by a scheduling algorithm. Once a file transfer is completed, the recipient will reply with a COMPLETE command message. If pending transfers are stored in the transferQueue, then this process will be repeated upon receiving this COMPLETE message by calling nextTransfer again, as shown in Figure~\ref{fig:file_transfer}.

The sendFile method accepts a Message and an Endpoint. A file payload will be created with the video file information from the Message. A byte payload will also be created using the ANALYSE command to indicate that this video should be analysed, the filename of the selected video file so that the download file will have the proper name, as well as the ID number of the video's file payload. Both of these payloads are then sent to the device represented by the endpoint parameter's ID field.

When a payload is received it is passed to the PayloadCallback's onPayloadReceived method. onPayloadReceived first checks if the payload is a byte payload or a file payload. All file payloads are treated the exact same way when they are received and start downloading, they are stored in the incomingFilePayloads dictionary for later retrieval. Whenever new file payload data is received, a call is made to onPayloadTransferUpdate to check if a file download has completed. This is necessary since a whole file typically cannot fit within a single packet. When a file download has completed, its payload is moved from the incomingFilePayloads dictionary to the completedFilePayloads dictionary, and its ID is passed to processFilePayload.

When a byte payload is received, its colon delimiters are used to split it into a String array. The first item in this array is the command which is passed through a switch statement to take the appropriate action. For ANALYSE and RETURN commands, the filename is stored in the filePayloadFilenames dictionary while the command itself is stored in the filePayloadCommands dictionary. Afterwards, the payload ID is passed to the processFilePayload method. The RETURN command also updates the UI by removing the sent video from the processing list and adding the returned result file to the results list to show that it has been successfully analysed. Upon receiving COMPLETE commands the UI is updated. The video is moved from the raw footage list to the processing list to indicate that the video was successfully transferred and is currently being processed by the remote device, nextTransfer is also called to start transferring the next video file

File payloads and byte payloads can arrive in any order. To address this, when payloads finish downloading, their IDs are passed to processFilePayload. This method first checks that both the file and byte payload for the given ID have successfully downloaded. If they have downloaded, a COMPLETE command message is sent to the original sender via sendCommandMessage and the video is ready for processing. The downloaded file has a generic name, so it is renamed to the filename retrieved from the filePayloadFilenames dictionary. The command retrieved from the filePayloadCommands dictionary determines what should be done with the downloaded file. For RETURN commands, the result file is simply moved to the results directory. The UI is also updated by removing the corresponding video item from the processing list and adding the result to the results list. For ANALYSE commands, the video is passed to VideoAnalysis for processing. Upon completion, the the results file along with the RETURN command is passed to sendFile so that it may delivered back to the device that sent it.

\begin{figure}[htb]
    \centering
    \includegraphics[width=0.53\textwidth]{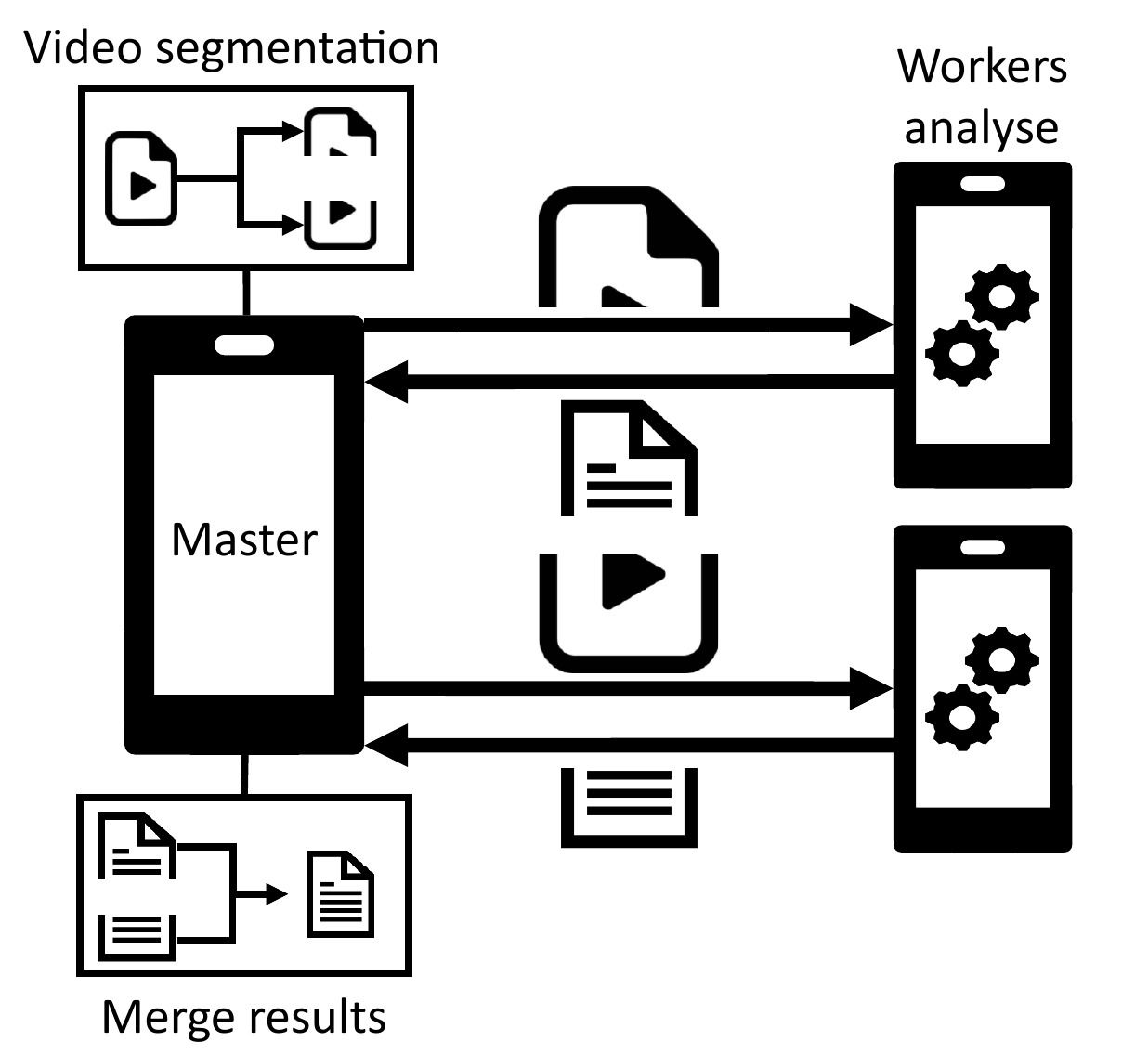}
    \caption{Analysis of a video split into two segments}
    \label{fig:segmentation}
\end{figure}

\subsubsection{Segmented Analysis}\label{sec:segmented_analysis}
Videos can be split into smaller segments by the master to ensure that there are multiple videos to be scheduled within a short time frame. The number of segments that a video is split into can be set by the user using a preference option or it may be automatically determined. A simple example of analysis where a video is split into two segments is demonstrated in Figure~\ref{fig:segmentation}. Before a video is added to the transferQueue, they are first split into equal segments with splitVideo. splitVideo will calculate the segment duration which it will use with FFmpeg's segment tool to split the video. The split video files are then stored in the segment directory and are named after the original video with a numerical suffix indicating its order. These segmented videos are then transmitted to worker devices where they are analysed like typical videos and return the results to the master. When the master receives the results for all video segments, it will use mergeResults to combine the segment result files into a single result file. This results file will be added to the result list.

\subsection{Scheduling Algorithm}\label{sec:algorithm}
Upon initiation of automatic downloading, the master will start to download videos in pairs, waiting a period of time equal to the video length before starting the next pair of downloads. Upon download completion, the master will make a decision based on the circumstances.

When the master is not connected to any other devices, it can only process videos by itself. No networking operations are performed, the master simply performs all video analysis tasks locally.

\begin{figure}[hb]
    \centering
    \includegraphics[width=0.71\textwidth]{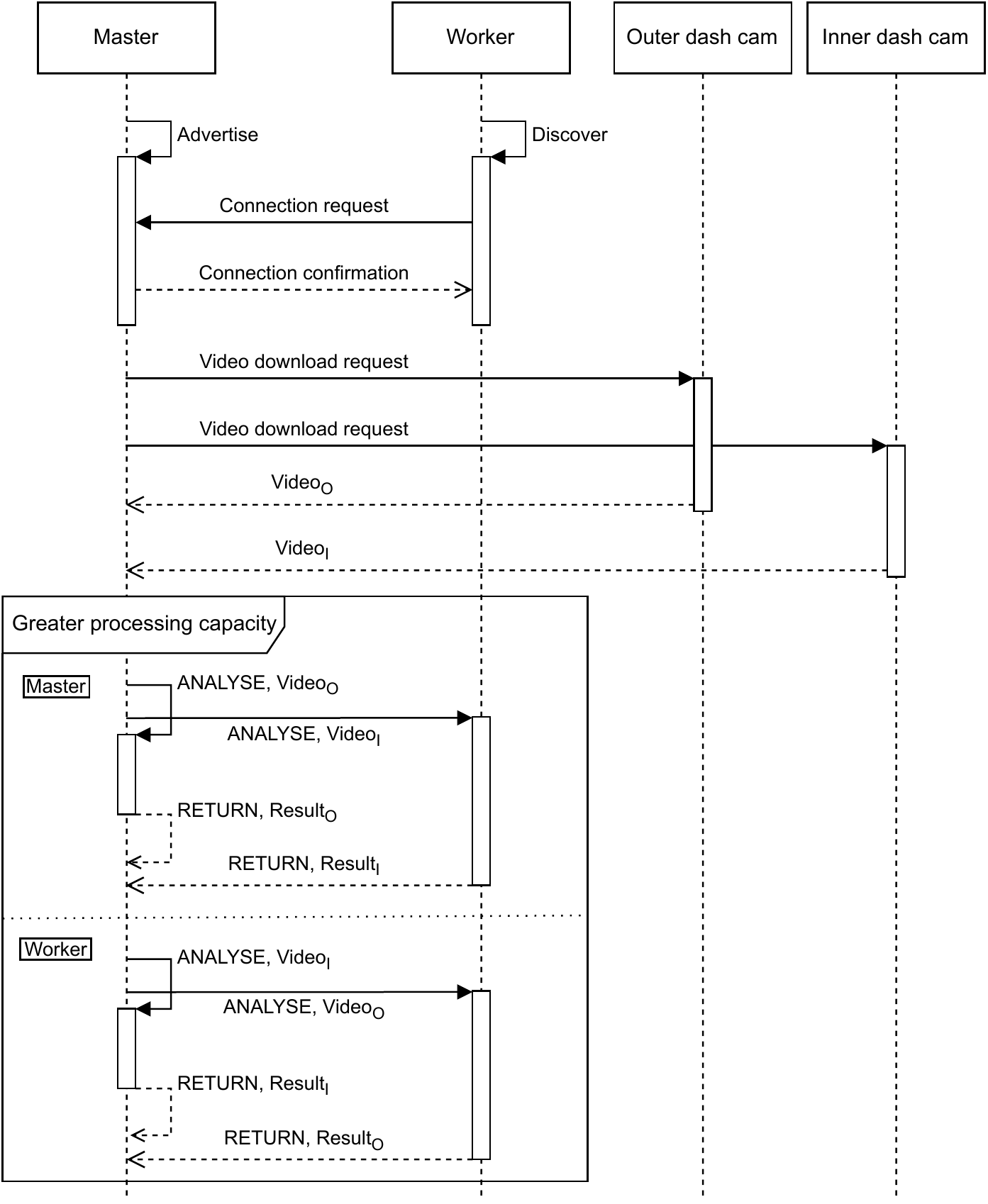}
    \caption{Algorithm process with two devices}
    \label{fig:algorithm_2n}
\end{figure}

If the master is connected to a single worker, then it will compare the processing capacities of the devices. If the master has greater processing capacity, then it will locally process outer videos and send the inner videos to the worker. The opposite will occur if the worker has grater processing capacity, with the master locally processing inner videos while the worker processes outer videos. This process is demonstrated in Figure~\ref{fig:algorithm_2n}.

If the master is connected to more than one worker and segmentation is disabled, then it will compare the processing capacities of the devices. If the master has greater processing capacity than all workers, then it will assign videos to itself for local processing if it is not already occupied with processing a video. If the master does not have the greatest processing capacity, then it will prioritise sending videos to workers instead, only processing videos locally if the workers are currently busy processing videos. The master decides which worker to assign a video to based on an algorithm that selects the unoccupied worker with the greatest processing capacity. If the master and all workers are busy, then the master will assign the video to the worker with the greatest processing capacity and shortest job queue.

If the master is connected to more than one worker and segmentation is enabled, then it will first assign the outer video to the device with the greatest processing capacity. It will then split the inner video into two segments of equal length, assigning them to the remaining devices.

\chapter{Evaluation}\label{chap:evalutation}
EDA was evaluated through a series of tests with differing network configurations. These tests demonstrate that EDA is capable of performing video analytics tasks on dash cam video in near real-time. However, significant optimisations were necessary in order to achieve this.

\section{Experimental Settings}\label{sec:experimental_settings}
The tests were performed with four Android smartphones and the VIOFO A129 dash cam. The VIOFO A129 was chosen for its inclusion of two cameras, reliability, and it offering an interface compatible with Android smartphones. The smartphones consisted of a Google Pixel~3, a Google Pixel~6, an OPPO Find X2 Pro, and a OnePlus~8. These smartphones are representative of Android devices with low, medium, and high processing capacities. Relevant hardware information of these devices is detailed in Table~\ref{tab:hardware}. All of the devices have 8-core CPUs, though the clock speeds of these CPUs as well as their RAM capacity differ from each other. The Pixel~3 has the lowest processing capacity with four 2.5Ghz cores, four 1.6Ghz cores, and 4GB of RAM. The Pixel~6 was found to have middling performance despite having two 2.8GHz cores, two 2.25GHz cores, four 1.8GHz cores and 8GB of RAM. The OnePlus~8 and Find X2 Pro both have high processing capacities with identical CPUs of one 2.84GHz core, three 2.42GHz cores, and four 1.8GHz cores. However, while the OnePlus~8 has 8GB of RAM, the Find X2 Pro has a greater amount at 12GB of RAM.

\begin{table}[hb]
    \centering
    \resizebox{\textwidth}{!}{%
        \begin{tabular}{l|l|l|l|l|l}
            Device                                                                     & CPU (\#cores $\times$ Ghz)                     & RAM (GB) & Battery        & Android & Processing \\
                                                                                       &                                                &          & Capacity (mAh) & Version & Capacity   \\ \hline
            Google Pixel 3~\cite{googlePixelPhoneHardware2022}                         & 4$\times$2.5 \& 4$\times$1.6                   & 4        & 2915           & 12      & Low        \\ \hline
            Google Pixel 6~\cite{googlePixelPhoneHardware2022,tibkenGoogleBuiltAI2021} & 2$\times$2.8 \& 2$\times$2.25 \& 4$\times$1.8  & 8        & 4614           & 12      & Medium     \\ \hline
            OnePlus 8~\cite{oneplusOnePlusSpecs2020,kressin2019Snapdragon8652019}      & 1$\times$2.84 \& 3$\times$2.42 \& 4$\times$1.8 & 8        & 4300           & 11      & High       \\ \hline
            OPPO Find X2 Pro~\cite{oppoOPPOFindX22022,kressin2019Snapdragon8652019}    & 1$\times$2.84 \& 3$\times$2.42 \& 4$\times$1.8 & 12       & 4260           & 10      & High
        \end{tabular}%
    }
    \caption{Hardware details and processing capacities of evaluated devices. Each mobile device has two or more sets of heterogeneous CPU cores, e.g., 4$\times$2.5Ghz cores and 4$\times$1.6 Ghz cores for the Pixel~3.}
    \label{tab:hardware}
\end{table}

\clearpage
The videos used to evaluate EDA are taken from the BDD100K dash cam video dataset~\cite{yuBDD100KDiverseDriving2020} and the DMD driver monitoring dataset~\cite{ortegaDMDLargeScaleMultimodal2020}. Videos from BDD100K are recorded from forward-facing dash cams, capturing urban environments in varying levels of traffic. Videos from DMD use inward-facing cameras to record a participant driving normally, or performing distracted actions such as talking on the phone or eating while in a parked (or simulated) car. Videos from both datasets have a resolution of 1280$\times$720 and a frame rate of 30 FPS, however, they have dissimilar lengths. BDD100K videos are all 40 seconds long, while the DMD videos vary in length from 1 to 9 minutes.
To ensure consistency between these datasets and to closer achieve real-time processing, the videos were split into segments of equal length. Two segmented sets were created, one consisted of 1600 segments of one-second length, while the other had 800 segments of two-second length. Half of the videos from these segmented sets were taken from BDD100K, the other half were taken from DMD.
The exact videos that were selected from these datasets, and their relation to their segments used in testing, are described in the \verb|filename_mapping.txt| file\footnote{Available in GitHub repository: \url{https://github.com/JaydenKing32/EdgeDashAnalytics}}. For the purposes of this project, BDD100K videos are referred to as outer videos, while DMD videos are referred to as inner videos. During testing, videos were downloaded as inner-outer pairs in emulation of them just being recorded from inward and outward-facing cameras.

Tests were performed with the one-second and two-second video segments. In the one-second tests, the one-second videos were loaded onto the master device beforehand and an artificial delay was added to simulate downloading time. The simulated downloading time was set to 350ms, half the time it took to download a two-second video from the dash cam, averaged between all devices. In the two-second tests, videos loaded onto the dash cam beforehand are downloaded by the master device via Wi-Fi. For both tests, the master waits a period of time equal to the test type's video length before initiating another pair of downloads, emulating the time taken for the dash cam to record the videos in a live setting. The reason why one-second video downloads are simulated are due to a inherent overhead delay of around 500ms between enqueuing a download and the download actually starting. This delay makes it impossible to download one-second videos in a real-time manner, as they cannot download faster than they are enqueued.

The setup of one-second and two-second tests slightly differ from one another. With one-second tests, test videos are pre-loaded onto the master device, the test video count is set to 800 inner-outer pairs, the wait period between downloads is set to one second, and simulated downloading is enabled with a 350ms delay. With two-second tests, the test video count is set to 400 inner-outer pairs, the wait period between downloads is set to two seconds, and the master device is connected to the dash cam via Wi-Fi. All other aspects of running one-second and two-second tests are identical to each other. This starts with the master connecting with workers with Nearby Connections, initiating downloads, collecting logs upon completion, and then wiping test-related files to prepare for the next test.

Testing comprised of multiple configurations of the test devices, these test devices being the Google Pixel~3, Google Pixel~6, OPPO Find X2 Pro, and OnePlus~8. One-node tests involved a single device processing all videos from the dataset. Two-node tests included the follow configurations: two strong devices, Find X2 Pro and OnePlus~8; a strong and a mid-range device, Find X2 Pro and Pixel~6; a mid-range and a weak device, Pixel~6 and Pixel~3. Three-node tests included the follow configurations: two strong and one mid-range device, Find X2 Pro, OnePlus~8, and Pixel~6; one strong, one mid-range, and one weak device, Find X2 Pro, Pixel~6, and Pixel~3.

\section{Results}\label{sec:results}
The results of each test run on EDA, in addition to an explanation of how the results were collected are presented below. The specific metrics used to evaluate EDA are the average video turnaround times, skip rates, and energy consumption. In particular, the time taken to complete various tasks and the energy consumed to do so is recorded for each test run. Times are recorded through the use of the Android logging tool, Logcat. All of the time values presented here are per-video averages. Additionally, each device's ESD value and the proportion of frames that are discarded due to early-stopping, termed the ``skip rate'', are included below.

\subsection{Metric Collection}\label{sec:metric_collection}
The recorded time values are split into six types: download, transfer, return, processing, wait, turnaround, and overhead. These values are all measured in milliseconds (ms). Download time is simply the duration of time between a video starting to download, and the video download finishing. A download time of 350ms was simulated for the one-second tests, while the actual time it took to download a video from the dash cam was recorded in the two-second tests. Transfer time is the time taken to transfer a video from the master device to a worker. Return time is the time taken to transfer result files from workers to the master. Processing time is the time taken to extract a video's frames, analyse said frames, and write the results to a JSON file. Wait time is the time between a video being received by a device and when the video starts processing. Wait time is primarily caused by a video waiting for another video to finish processing, but it can also be caused by system delays that occur after a worker receives a video. Turnaround is the duration of time between a video starting to download and that video's result file being received by the master. If a video's turnaround is shorter than the length of the video itself, then it can be said that it was processed in a near real-time manner. Finally, overhead represents delays caused by the OS or libraries that are not accounted for by the other time value types. The largest contributor to overhead is the delay in between starting a file transfer and the transfer actually starting, though it has other contributing factors such as process start-up delays. Overhead is calculated by taking the sum of download, transfer, return, processing, and wait times, and deducting it from the turnaround time.

Recording energy consumption is necessary due to the devices being battery powered, as the amount of time they can operate is limited. While vehicles often offer power outlets for charging, they may not have enough for all devices. We use two methods for recording energy consumption. The first method is based on the approach by \textcite{silvaEnergyawareAdaptiveOffloading2021}, taking the power readings provided by the Android API. These power values are measured in milliwatts (mW) and are presented as per-video averages. The second method concerns battery usage and involves recording device battery levels at the start and at the end of test runs. The difference between these values gives the amount of battery power consumed as a percentage of total battery capacity. While using specialised hardware to physically monitor a device's energy consumption may be more accurate, the use of the Android API was found to be sufficient.

We use early-stopping for some devices to ensure near real-time turnarounds.
The early-stop divisor (ESD) value determines when video analysis ends early. Essentially, higher ESD values will result in video analysis stopping earlier, leading to more video frames being discarded.
Weaker devices such as the Pixel~3 and Pixel~6 require the use of high ESD values in order to reach near real-time turnarounds, while stronger devices like the Find X2 Pro and OnePlus~8 only need low ESD values.

\begin{figure}[!hb]
    \centering
    \includegraphics[width=0.7\textwidth]{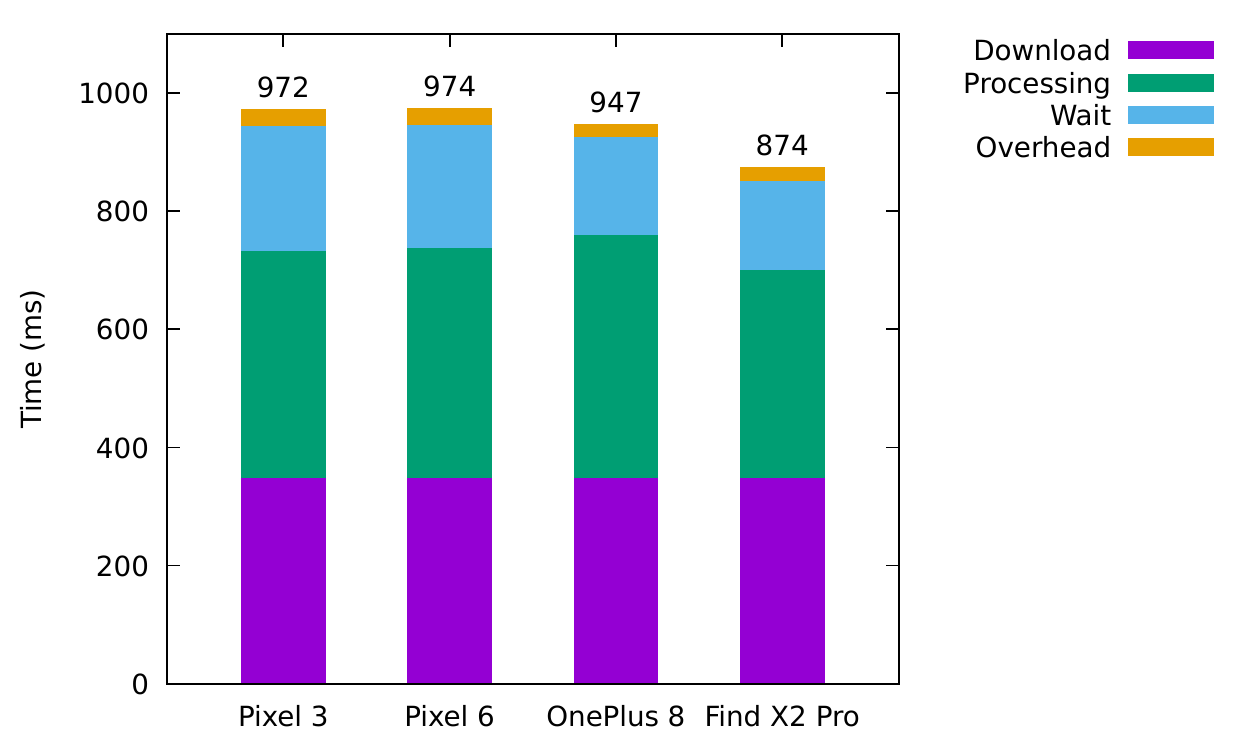}
    \caption{Average time taken by tasks in one-second one-node tests. Values add up to average turnaround.}
    \label{fig:results_1s_1n}
\end{figure}

\begin{table}[!hb]
    \centering
    \resizebox{\textwidth}{!}{%
        \begin{tabular}{l|l|l|l|l|l|l}
            Device      & Processing (ms) & Wait (ms) & Overhead (ms) & Turnaround (ms) & ESD & Skip rate \\ \hline
            Pixel 3     & 385             & 211       & 26            & 972             & 2.8 & 59.2\%    \\ \hline
            Pixel 6     & 389             & 208       & 27            & 974             & 2.6 & 14.5\%    \\ \hline
            OnePlus 8   & 411             & 166       & 20            & 947             & 0   & 0\%       \\ \hline
            Find X2 Pro & 352             & 150       & 22            & 874             & 0   & 0\%
        \end{tabular}%
    }
    \caption{One-second one-node test results, simulated download time of 350ms.}
    \label{tab:results_1s_1n}
\end{table}

\subsection{Turnaround and Skip Rate}\label{sec:turnaround_and_skip_rate}
Here, we present the average video turnaround times and skip rates with respect to two video time granularities. In particular, near real-time turnaround times are significantly enabled through the use of early stopping when the resource capacity of participating devices are not sufficient to process videos in a real-time manner.

\begin{figure}[ht]
    \centering
    \includegraphics[width=0.8\textwidth]{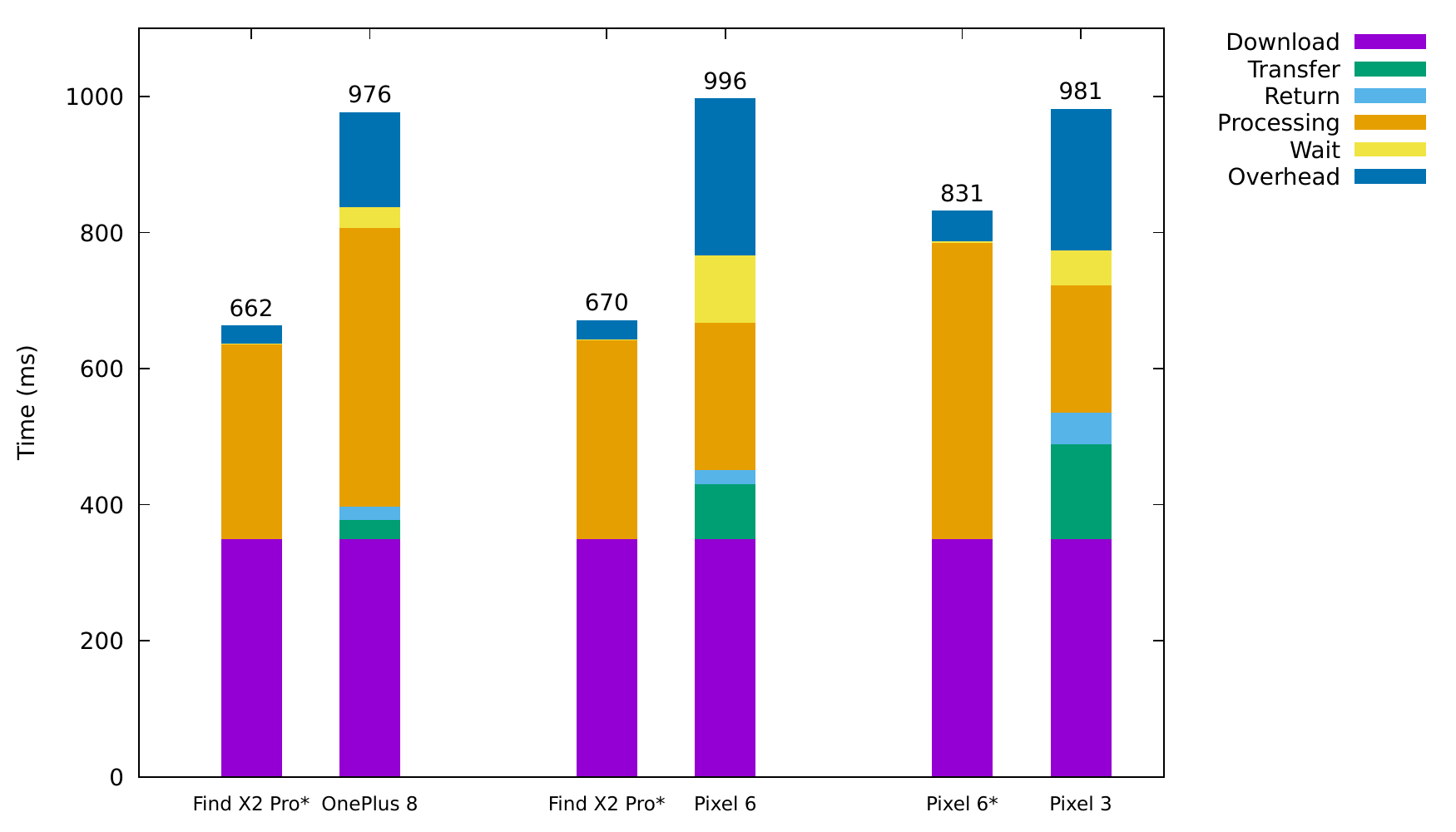}
    \caption{Average time taken by tasks in one-second two-node tests. Values add up to average turnaround. * indicate master device.}
    \label{fig:results_1s_2n}
\end{figure}

\begin{table}[ht]
    \centering
    \resizebox{\textwidth}{!}{%
        \begin{tabular}{l|l|l|l|l|l|l|l|l}
            Device       & Transfer (ms) & Return (ms) & Processing (ms) & Wait (ms) & Overhead (ms) & Turnaround (ms) & ESD & Skip rate \\ \hline
            Find X2 Pro* & n/a           & n/a         & 287             & 1         & 24            & 662             & 0   & 0\%       \\
            OnePlus 8    & 29            & 19          & 410             & 30        & 138           & 976             & 2.5 & 26.1\%    \\ \hline
            Find X2 Pro* & n/a           & n/a         & 293             & 1         & 26            & 670             & 0   & 0\%       \\
            Pixel 6      & 81            & 21          & 216             & 100       & 228           & 996             & 5   & 80.5\%    \\ \hline
            Pixel 6*     & n/a           & n/a         & 436             & 2         & 43            & 831             & 0   & 0\%       \\
            Pixel 3      & 140           & 46          & 187             & 52        & 206           & 981             & 6   & 98.7\%
        \end{tabular}%
    }
    \caption{One-second two-node test results, simulated download time of 350ms, * indicate master device.}
    \label{tab:results_1s_2n}
\end{table}

\subsubsection{One-second Tests}\label{sec:one_second_tests}
As seen in Figure~\ref{fig:results_1s_1n} and Table~\ref{tab:results_1s_1n}, displaying the results of the one-node tests, the OnePlus~8 and Find X2 Pro were able to achieve average turnarounds of 947ms and 874ms respectively, without using early-stopping. However, in order to reach near real-time turnaround, the Pixel~3 had to use an ESD of 4 resulting in a skip rate of 78\%, while the Pixel~6 used an ESD of 3 resulting in a skip rate of 31.7\%.

\begin{figure}[hb]
    \centering
    \includegraphics[width=0.8\textwidth]{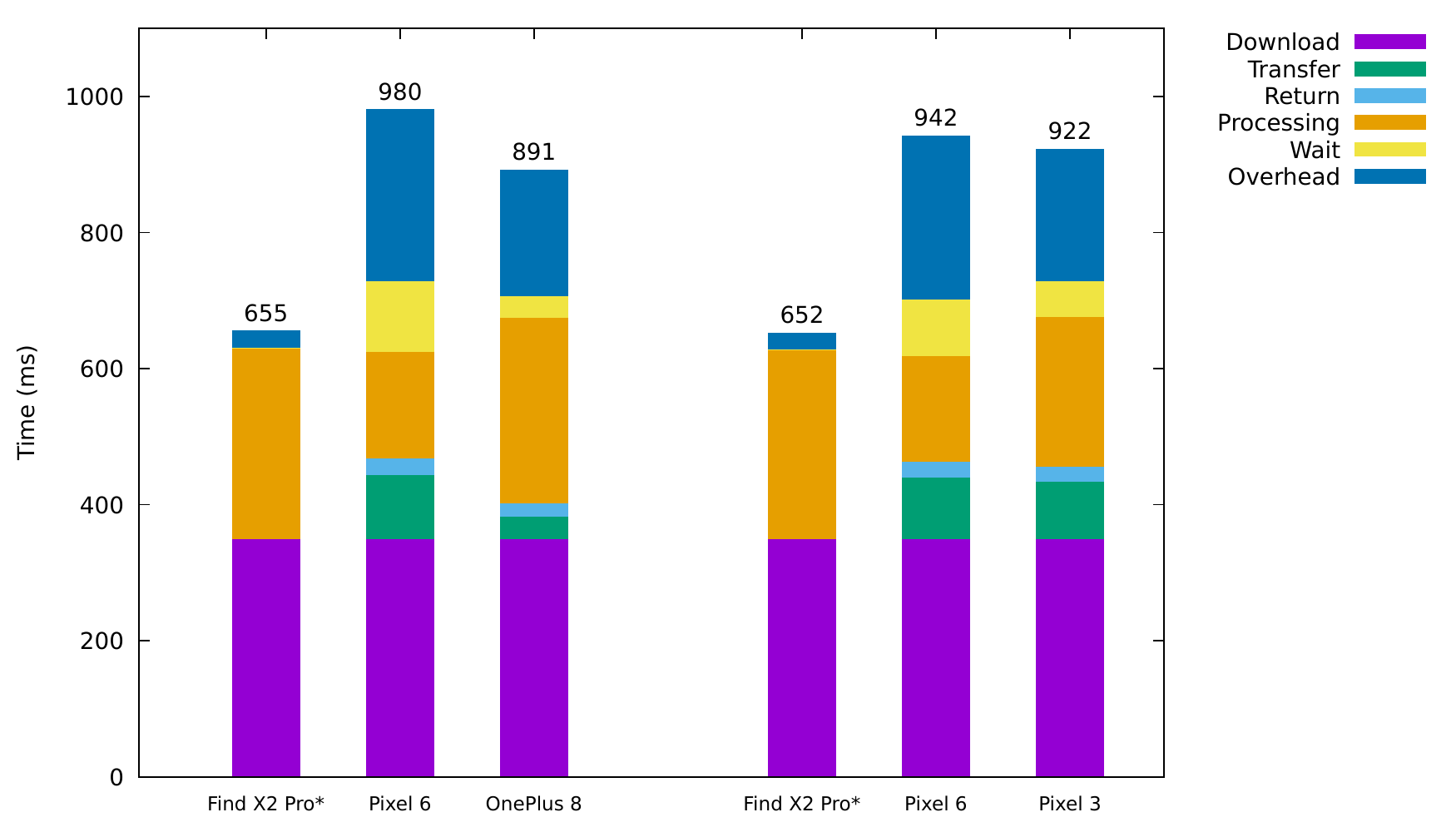}
    \caption{Average time taken by tasks in one-second three-node tests. Values add up to average turnaround. * indicate master device.}
    \label{fig:results_1s_3n}
\end{figure}

\begin{table}[hb]
    \centering
    \resizebox{\textwidth}{!}{%
        \begin{tabular}{l|l|l|l|l|l|l|l|l}
            Device       & Transfer (ms) & Return (ms) & Processing (ms) & Wait (ms) & Overhead (ms) & Turnaround (ms) & ESD & Skip rate \\ \hline
            Find X2 Pro* & n/a           & n/a         & 281             & 1         & 23            & 655             & 0   & 0\%       \\
            Pixel 6      & 95            & 24          & 156             & 104       & 251           & 980             & 4   & 90.9\%    \\
            OnePlus 8    & 33            & 20          & 273             & 32        & 183           & 891             & 0   & 0\%       \\ \hline
            Find X2 Pro* & n/a           & n/a         & 278             & 1         & 23            & 652             & 0   & 0\%       \\
            Pixel 6      & 91            & 23          & 155             & 84        & 239           & 942             & 4   & 89.2\%    \\
            Pixel 3      & 85            & 22          & 220             & 53        & 192           & 922             & 3   & 83\%
        \end{tabular}%
    }
    \caption{One second three-node test results, simulated download time of 350ms, * indicate master device.}
    \label{tab:results_1s_3n}
\end{table}

For the networked tests, all workers had to use higher ESD values to compensate for the extra time taken by network operations such as sending the videos to workers and returning the results to the master. On the other hand, the master devices did not use early-stopping at all, as they avoided the extra networking operations and only had to process outer videos, offloading the inner videos to the workers.

As seen in Figure~\ref{fig:results_1s_2n} and Table~\ref{tab:results_1s_2n}, displaying the results of the two-node tests, the run with the OnePlus~8 acting as a worker with the Find X2 Pro master was able to reach a turnaround of 976ms with an ESD of 2.5, resulting in a skip rate of 26.1\%. When the Pixel~6 acted as a worker with the Find X2 Pro master, it achieved a turnaround of 996ms with an ESD of 5, leading to a skip rate of 80.5\%. Finally, the Pixel~3 acting as a worker with the Pixel~6 master had a 981ms turnaround with an ESD of 6, resulting in a high skip rate of 98.7\%.

The three-node tests utilised segmentation, where the master splits inner videos into two equal halves. All three-node tests used the Find X2 Pro as master. The first test had the Pixel~6 and OnePlus~8 take on worker roles. As seen in Figure~\ref{fig:results_1s_3n} and Table~\ref{tab:results_1s_3n}, the OnePlus~8 was able to achieve a turnaround of 891ms without early-stopping due to it only needing to process a half-second segment. The Pixel~6 still needed to use an ESD of 3.5 to reach a turnaround of 980ms, leading to a skip rate of 90.9\%. The second test utilised the Pixel~3 and Pixel~6 as workers. Again, the Pixel~6 used an ESD of 3.5, reaching a turnaround of 942ms and a skip rate of 89.2\%. Oddly, the Pixel~3 was able to use a lower ESD of 2.5 to reach a turnaround of 922ms and a skip rate of 83\%. This is due to the Pixel~3 having slightly higher network speeds and lower overhead delays compared to to the Pixel~6, despite its analysis speeds being lower.
\clearpage

\begin{table}[ht]
    \centering
    \resizebox{\textwidth}{!}{%
        \begin{tabular}{l|l|l|l|l|l|l|l}
            Device      & Download (ms) & Processing (ms) & Wait (ms) & Overhead (ms) & Turnaround (ms) & ESD & Skip rate \\ \hline
            Pixel 3     & 893           & 766             & 259       & 34            & 1952            & 2.7 & 37.1\%    \\ \hline
            Pixel 6     & 759           & 783             & 354       & 29            & 1925            & 0   & 0\%       \\ \hline
            OnePlus 8   & 598           & 763             & 445       & 22            & 1828            & 0   & 0\%       \\ \hline
            Find X2 Pro & 613           & 649             & 359       & 23            & 1644            & 0   & 0\%
        \end{tabular}%
    }
    \caption{Two-second one-node test results.}
    \label{tab:results_2s_1n}
\end{table}

\begin{figure}[ht]
    \centering
    \includegraphics[width=0.8\textwidth]{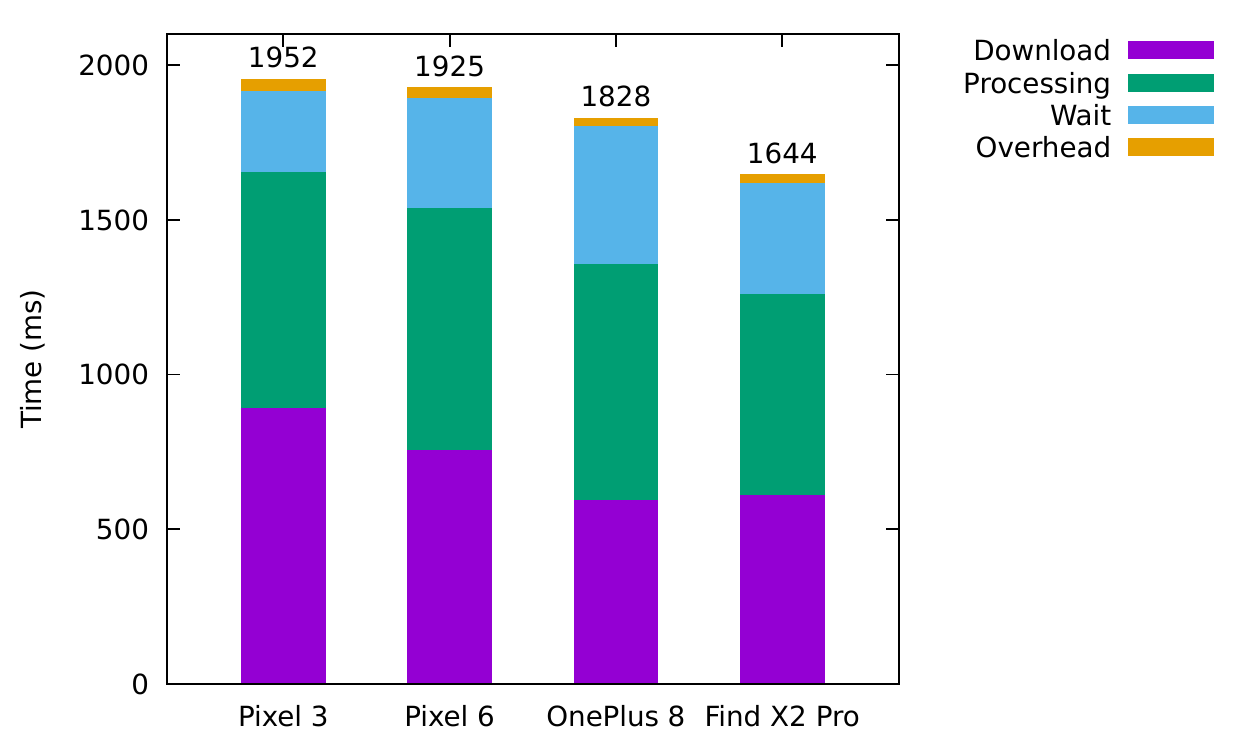}
    \caption{Average time taken by tasks in two-second one-node tests. Values add up to average turnaround.}
    \label{fig:results_2s_1n}
\end{figure}

\subsubsection{Two-second Tests}\label{sec:two_second_tests}
Unlike the simulated downloads in the one-second tests, the master device in two-second tests actually downloads video files from a dash cam. Despite the potential variance in download times, almost all two-second tests were able to reach shorter turnarounds with lower skip rates thanks to the reduced overhead of dealing with fewer and larger video files.
As seen in Figure~\ref{fig:results_2s_1n} and Table~\ref{tab:results_2s_1n}, showing the one-node test results, only the Pixel~3 needed to use early-stopping with an ESD of 3.5 in order to achieve a turnaround of 1710ms and a skip rate of 55.2\%.

\begin{figure}[!hb]
    \centering
    \includegraphics[width=0.7\textwidth]{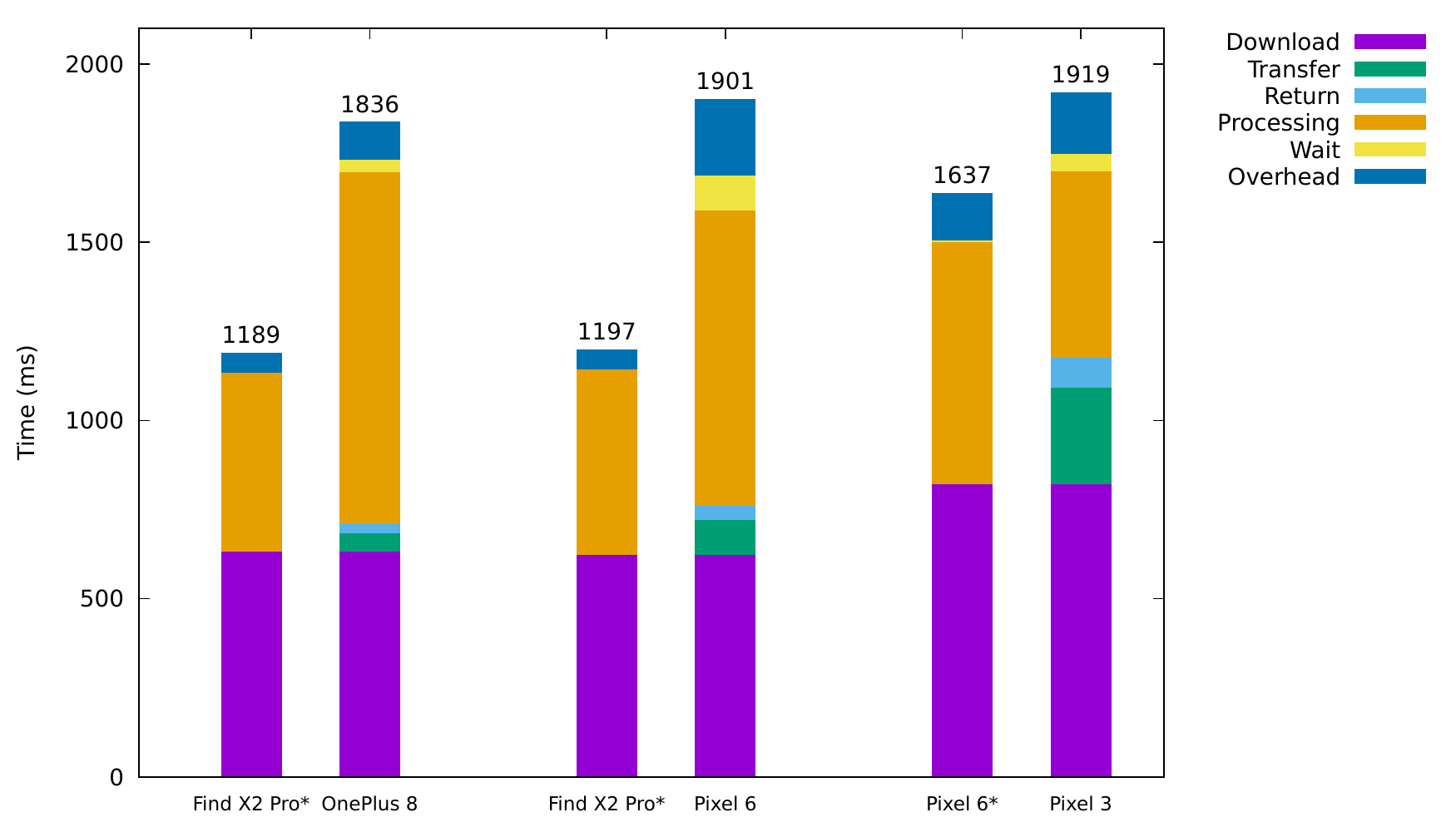}
    \caption{Average time taken by tasks in two-second two-node tests. Values add up to average turnaround.}
    \label{fig:results_2s_2n}
\end{figure}

\begin{table}[!hb]
    \centering
    \resizebox{\textwidth}{!}{%
        \begin{tabular}{l|l|l|l|l|l|l|l|l|l}
            Device       & Download (ms)        & Transfer (ms) & Return (ms) & Processing (ms) & Wait (ms) & Overhead (ms) & Turnaround (ms) & ESD & Skip rate \\ \hline
            Find X2 Pro* & \multirow{2}{*}{633} & n/a           & n/a         & 503             & 0         & 53            & 1189            & 0   & 0\%       \\
            OnePlus 8    &                      & 51            & 30          & 985             & 35        & 102           & 1836            & 0   & 0\%       \\ \hline
            Find X2 Pro* & \multirow{2}{*}{625} & n/a           & n/a         & 521             & 0         & 51            & 1197            & 0   & 0\%       \\
            Pixel 6      &                      & 98            & 39          & 830             & 96        & 213           & 1901            & 0   & 0\%       \\ \hline
            Pixel 6*     & \multirow{2}{*}{824} & n/a           & n/a         & 678             & 6         & 129           & 1637            & 3   & 12.7\%    \\
            Pixel 3      &                      & 270           & 84          & 523             & 48        & 170           & 1919            & 4   & 69.2\%
        \end{tabular}%
    }
    \caption{Two-second two-node test results, * indicate master device.}
    \label{tab:results_2s_2n}
\end{table}

Neither the two-node test with the Find X2 Pro as master, one with a OnePlus~8 worker and the other with a Pixel~6 worker, needed early-stopping to achieve near real-time turnaround, as seen in Figure~\ref{fig:results_2s_2n} and Table~\ref{tab:results_2s_2n}. However, the Pixel~6 and Pixel~3 master/worker pair needed to use early-stopping to compensate for slower transfer speeds. The Pixel~6 used an ESD of 3 to reach a turnaround of 1637ms with a skip rate of 12.7\%, while the Pixel~3 used an ESD of 4 to achieve a turnaround of 1919ms and a skip rate of 69.2\%.

\begin{figure}[!hb]
    \centering
    \includegraphics[width=0.7\textwidth]{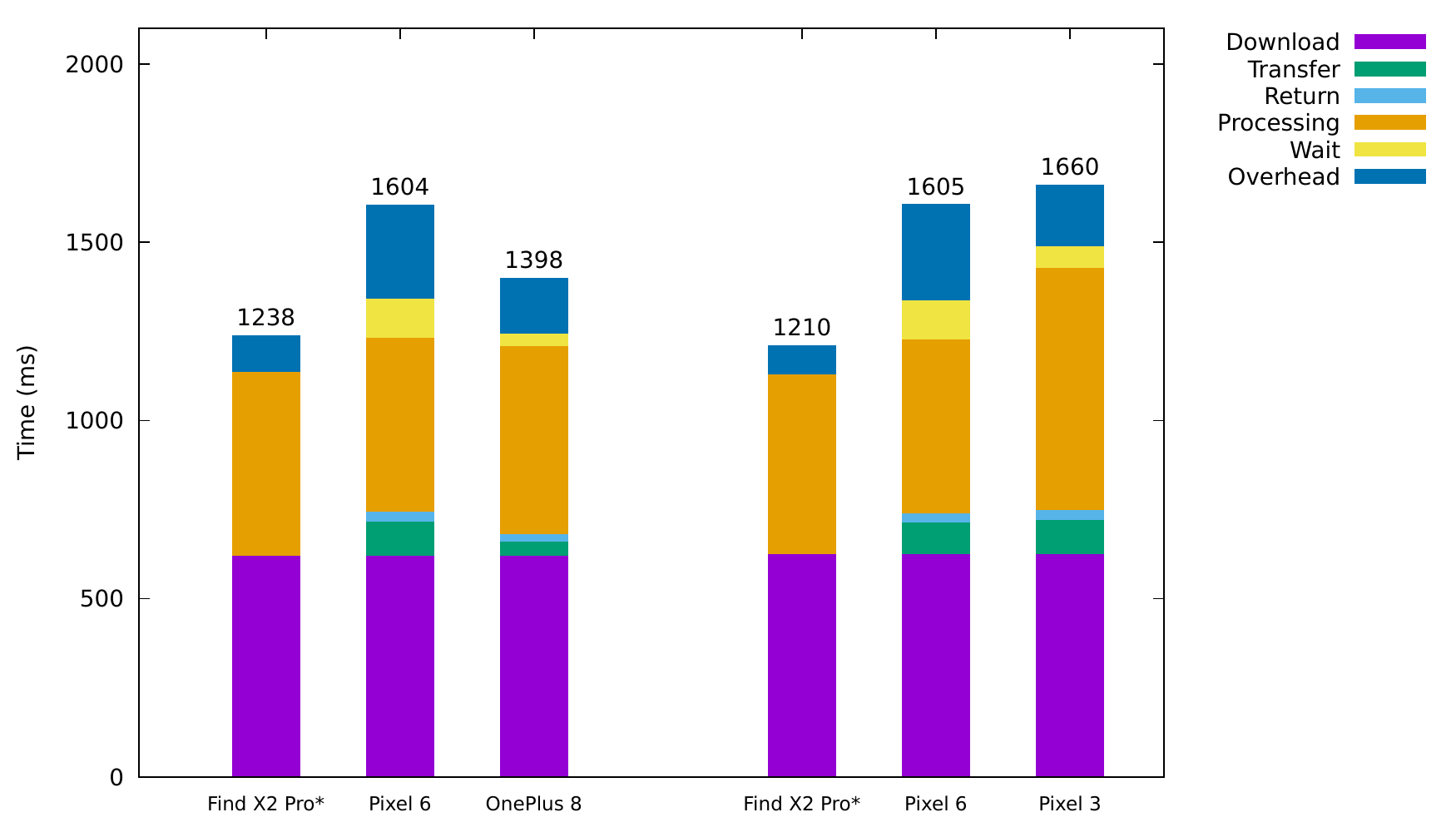}
    \caption{Average time taken by tasks in two-second three-node tests. Values add up to average turnaround.}
    \label{fig:results_2s_3n}
\end{figure}

\begin{table}[!ht]
    \centering
    \resizebox{\textwidth}{!}{%
        \begin{tabular}{l|l|l|l|l|l|l|l}
            Device       & Download (ms)        & Transfer (ms) & Return (ms) & Processing (ms) & Wait (ms) & Overhead (ms) & Turnaround (ms) \\ \hline
            Find X2 Pro* & \multirow{3}{*}{623} & n/a           & n/a         & 515             & 1         & 99            & 1238            \\
            Pixel 6      &                      & 96            & 27          & 488             & 110       & 260           & 1604            \\
            OnePlus 8    &                      & 38            & 21          & 528             & 35        & 153           & 1398            \\ \hline
            Find X2 Pro* & \multirow{3}{*}{626} & n/a           & n/a         & 506             & 0         & 78            & 1210            \\
            Pixel 6      &                      & 89            & 27          & 487             & 109       & 267           & 1605            \\
            Pixel 3      &                      & 96            & 28          & 680             & 60        & 170           & 1660
        \end{tabular}%
    }
    \caption{Two-second three-node test results, no use of early-stopping, * indicate master device.}
    \label{tab:results_2s_3n}
\end{table}

None of the two-second three-node tests required early-stopping to achieve near real-time turnarounds. The test using the Find X2 Pro as master with the Pixel~6 and OnePlus~8 as workers was able to achieve an average turnaround of 1413ms, as seen in Figure~\ref{fig:results_2s_3n} and Table~\ref{tab:results_2s_3n}. The test using the Find X2 Pro as master with Pixel~3 and Pixel~6 as workers was slightly slower, with an average turnaround of 1492ms.

\subsection{Energy Consumption}\label{sec:energy_consumption}
Overall, the energy consumption of participating devices is insignificant in the context of battery usage, only 1-8\% of total battery capacity was consumed to analyse 1600s of video data.

\begin{table}[htb]
    \centering
    \resizebox{0.9\textwidth}{!}{%
        \begin{tabular}{c|l|l|l|l}
            Node count         & Device       & Turnaround (ms) & Average power (mW) & Battery usage \\ \hline
            \multirow{4}{*}{1} & Pixel 3      & 972             & 19.175             & 8\%           \\ \cline{2-5}
                               & Pixel 6      & 974             & 35.935             & 5\%           \\ \cline{2-5}
                               & OnePlus 8    & 947             & 110.208            & 5\%           \\ \cline{2-5}
                               & Find X2 Pro  & 874             & 172.817            & 5\%           \\ \hline\hline
            \multirow{6}{*}{2} & Find X2 Pro* & 662             & 194.812            & 3\%           \\
                               & OnePlus 8    & 976             & 142.679            & 4\%           \\ \cline{2-5}
                               & Find X2 Pro* & 670             & 184.591            & 3\%           \\
                               & Pixel 6      & 996             & 7.566              & 0\%           \\ \cline{2-5}
                               & Pixel 6*     & 831             & 30.347             & 3\%           \\
                               & Pixel 3      & 981             & 14.994             & 3\%           \\ \hline\hline
            \multirow{6}{*}{3} & Find X2 Pro* & 655             & 217.565            & 3\%           \\
                               & Pixel 6      & 980             & 4.178              & 2\%           \\
                               & OnePlus 8    & 891             & 119.782            & 3\%           \\ \cline{2-5}
                               & Find X2 Pro* & 652             & 211.451            & 3\%           \\
                               & Pixel 6      & 942             & 7.332              & 1\%           \\
                               & Pixel 3      & 922             & 15.872             & 3\%
        \end{tabular}%
    }
    \caption{Turnaround and energy results of one second tests, * indicate master device.}
    \label{tab:energy_1s}
\end{table}

Looking at the energy consumption results in Table~\ref{tab:energy_1s} and Table~\ref{tab:energy_2s}, it is clear that the Find X2 Pro consistently uses much more energy than the OnePlus~8, while the Pixel~3 and Pixel~6 use much less energy. This is demonstrated in the single-node section of Table~\ref{tab:energy_1s}'s one-second test results, with the Pixel~3 having the lowest average power consumption of 19.175mW, followed by the Pixel~6 having a slightly greater average power consumption of 35.935mW, followed by a large jump to the OnePlus~8 at 110.208mW, ending with another large jump to the Find X2 Pro at 172.817mW. This trend is somewhat disrupted with the two-second one-node tests, where the Pixel~3's average power consumption of 96.031mW is higher than the Pixel~6's 57.537mW consumption. However, the other devices are consistent with the trend with the OnePlus~8's much greater power consumption of 217.6mW, followed by the even larger Find X2 Pro consumption of 353.838mW.

\begin{table}[htb]
    \centering
    \resizebox{0.9\textwidth}{!}{%
        \begin{tabular}{c|l|l|l|l}
            Node count         & Device       & Turnaround (ms) & Average power (mW) & Battery usage \\ \hline
            \multirow{4}{*}{1} & Pixel 3      & 1952            & 96.031             & 8\%           \\ \cline{2-5}
                               & Pixel 6      & 1925            & 57.537             & 5\%           \\ \cline{2-5}
                               & OnePlus 8    & 1828            & 217.600            & 4\%           \\ \cline{2-5}
                               & Find X2 Pro  & 1644            & 353.838            & 4\%           \\ \hline\hline
            \multirow{6}{*}{2} & Find X2 Pro* & 1189            & 423.257            & 3\%           \\
                               & OnePlus 8    & 1836            & 286.884            & 4\%           \\ \cline{2-5}
                               & Find X2 Pro* & 1197            & 432.826            & 3\%           \\
                               & Pixel 6      & 1901            & 19.140             & 1\%           \\ \cline{2-5}
                               & Pixel 6*     & 1637            & 32.368             & 3\%           \\
                               & Pixel 3      & 1919            & 30.154             & 4\%           \\ \hline\hline
            \multirow{6}{*}{3} & Find X2 Pro* & 1238            & 431.849            & 3\%           \\
                               & Pixel 6      & 1604            & 11.589             & 3\%           \\
                               & OnePlus 8    & 1398            & 237.435            & 3\%           \\ \cline{2-5}
                               & Find X2 Pro* & 1210            & 430.991            & 4\%           \\
                               & Pixel 6      & 1605            & 14.921             & 2\%           \\
                               & Pixel 3      & 1660            & 80.790             & 5\%
        \end{tabular}%
    }
    \caption{Turnaround and energy results of two second tests, * indicate master device.}
    \label{tab:energy_2s}
\end{table}

At first it would appear as though there is a strong correlation between energy use and processing capacity. However, there is a discrepancy in regards to the networked tests, where the Pixel~3 consumes more power than the Pixel~6, despite it having less processing capacity. Table~\ref{tab:energy_1s} demonstrates that when acting as workers, the Pixel~3 consumed almost double the average power consumption of the Pixel~6 for the one-second two-node tests. This difference is slightly higher with the one-second three-node tests, where the Pixel~3 consumed just over double the Pixel~6's average power consumption. This trend continues with the two-second tests shown by Table~\ref{tab:energy_2s}, where the Pixel~3's single-node test consumed 38mW per-video more than the Pixel~6. The two-second two-node tests present a smaller difference, when acting as workers, the Pixel~3 consumed 11mW per-video more than the Pixel~6. Finally, there was a much greater gap in the two-second three-node tests, where the Pixel~3 consumed over 65mW per-video more than the Pixel~6. This discrepancy of the weaker Pixel~3 consuming more power than the computationally stronger Pixel~6 could be attributed to power inefficiencies of the Pixel~3 that were corrected later in the Pixel~6. Alternatively, the recorded power values may have been influenced by inaccuracies present in Android's battery API. However, the use of specialised hardware would be required to verify these theories with any degree of certainty.

Despite the large differences in average power consumption, all of the devices only used 5\% of their total battery capacity for one-second single-node tests, except for the Pixel~3 which used 8\%, as shown in Table~\ref{tab:energy_1s}. Table~\ref{tab:energy_2s} demonstrates similar levels of battery consumption for the two-second single-node tests, where the Pixel~3 used 8\%, the Pixel~6 used 5\%, and both the OnePlus~8 and the Find X2 Pro used 4\%. The inconsistency between highly varied per-video power consumption and similar battery usage values could be explained by different battery capacities of the devices, as shown in Table~\ref{tab:hardware}.

\chapter{Conclusion}\label{chap:conclusion}
In this thesis, we have addressed issues of near real-time video analytics with dash cams and resource-constrained mobile devices. We have developed EdgeDashAnalytics (EDA) as a solution system that incorporates several optimisation techniques, for detecting road hazards and driver distractedness. Clearly, driver safety can be improved through near real-time video analytics that are achieved with mobile devices and dash cams. However, there are challenges present in reaching such a goal. A significant challenge is latency, results must be produced with low latency in order for video analytics to be effective. Cloud computing is unsuitable due to the high latency and bandwidth consumption involved in transmitting video data over long distances. Edge computing is a viable solution for dash cam video analytics as it makes it possible to avoid high latency and bandwidth consumption. With data being processed close to its source, data transmission should have a very small impact on latency. Although EDA has room to improve, it has demonstrated the feasibility and potential for distributed edge-based real-time video analytics on the move. These claims have been proven by experiment results that showed low latency and energy consumption.

\chapter{Future Work}\label{chap:future_work}
While EDA has shown great potential for real-time video analytics on the move with dash cams and resource-constrained mobile devices, there are a number of ways that it can be improved. We plan to improve EDA in areas such as: dynamic adjustment of the ESD value, modifying the amount of scaling applied to video frames, adding temporal analysis, using specifically tuned TensorFlow models, and switching to alternate forms of video data.

The most obvious improvement that can be made to EDA is the implementation of dynamic ESD adjustment. Currently, the ESD value is manually set prior to executing a test run. Instead, the ESD value should be automatically raised when a video's turnaround exceeds its length, eventually leading to an ESD value that results in real-time turnaround. This would be fairly easy to implement for master devices only, as the master device is responsible for recording turnaround times. However, this would be difficult to implement for workers as they are currently unaware of turnaround times. One potential solution would be to include the previous video's turnaround time along with the video messages sent to workers so that they can calculate ESD adjustments by themselves. Alternatively, the master could be responsible for adjusting worker ESD's. The master would calculate new ESD values based on worker turnaround times and instruct workers to change their ESD when necessary through a new type of message. There are several important questions regarding the implementation of dynamic ESD adjustment:

\begin{itemize}
    \item How much should the ESD be adjusted by? It would be simple to increment ESD by a constant value, however, this may introduce inefficiencies. If the adjustment value is too high then ESD could overshoot its ideal value, if its too low then it could take a long time until the ideal ESD value is reached. Alternatively, the ESD could be adjusted by a value that is proportional to the difference in turnaround time and video length.
    \item Should the ESD only be increased, or should it be decreased as well? The turnaround of videos is not constant, it can vary due to the inconsistency of some processes such as download times. If ESD is always raised whenever a video's turnaround exceeds its length, then infrequently high turnarounds may result in the ESD being set might higher than its ideal value. Instead, whenever a video's turnaround is lower than its length by some threshold, the ESD value could be lowered. However, this adjustment should not be too sensitive, otherwise the ESD may fluctuate wildly throughout execution of many videos.
    \clearpage
    \item What should occur when turnaround cannot be reduced below a video's length? This can be caused by situations such as the use of a device with low computational resources, or interference resulting in very slow download times. It would not make sense to constantly increase ESD even after it's high enough that all frames are skipped and turnaround is not reduced any further. A procedure could be added whereby ESD adjustments are halted when ESD reaches the point that all frames in a video are skipped. However, what comes after this is another issues. One solution could be that EDA recognises that it is not capable of processing any frames in a real-time manner and so it halts its operation and may display some kind of error message. Alternatively, it could reduce ESD so that some frames are still processed even with high turnarounds.
\end{itemize}

Further testing should be performed with the amount of scaling applied to video frames prior to passing them on for analysis. Downscaling frames was one of the optimisations made to reduce processing times, however, this reduces the accuracy of results. Currently, frames are downscaled to match the width of the TensorFlow model that will process it. A potential improvement could be adjusting the amount of scaling based on a device's processing capacity.

A significant improvement could be made through the addition of temporal analysis. Currently each frame is processed individually without any consideration of prior frames. With temporal analysis, the context of prior frames is used in analysis, improving detection results.

Only general-purpose TensorFlow models such as MobileNet and MoveNet have been used thus far. Detection accuracy could be greatly improved if models that were tailored to the task were used instead.

The use of alternative forms of video data instead of video files should be investigated. A lot of processing time is wasted by extracting frames from video files, this time could be saved if the data was transmitted in a form that could be immediately analysed. It was necessary to use video files due to the tested dash cams lacking a proper API that allows data streaming. This may be solved by finding a dash cam or similar device that does offer such an API. Alternatively, could just use a mobile device in place of the dash cam.

\clearpage
\emergencystretch=1em
\pdfbookmark{References}{sec:references}\printbibliography

@misc{allianzRiseDashCam2019,
  title    = {The Rise of the Dash Cam: {{New}} Data Reveals 1 in 5 {{Aussies}} Are Using Dash Cams on Our Roads},
  author   = {Allianz},
  year     = {2019},
  month    = sep,
  url      = {https://www.allianz.com.au/media/news/2019/the-rise-of-the-dash-cam-new-data-reveals-1-in-5-aussies-are-using-dash-cams-on-our-roads}
}

@inproceedings{besencziFrameworkDistributedProcessing2014,
  title     = {A Framework for Distributed Processing on an Offline Cell Phone Network},
  author    = {Besenczi, Ren{\'a}t{\'o} and Szitha, Krist{\'o}f and Hajdu, Andr{\'a}s},
  year      = {2014},
  month     = nov,
  booktitle = {2014 5th {{IEEE Conference}} on {{Cognitive Infocommunications}} ({{CogInfoCom}})},
  publisher = {{IEEE}},
  pages     = {257--262},
  doi       = {10.1109/CogInfoCom.2014.7020457}
}

@inproceedings{dangeSchedulingTaskCollaborative2016,
  title     = {Scheduling of Task in Collaborative Environment Using Mobile Cloud},
  author    = {Dange, Nikhil and Devadkar, Kailas and Kalbande, Dhananjay},
  year      = {2016},
  month     = dec,
  booktitle = {2016 {{International Conference}} on {{Global Trends}} in {{Signal Processing}}, {{Information Computing}} and {{Communication}} ({{ICGTSPICC}})},
  publisher = {{IEEE}},
  pages     = {579--583},
  doi       = {10.1109/ICGTSPICC.2016.7955367}
}

@inproceedings{fernandoMobileCrowdComputing2012,
  title     = {Mobile {{Crowd Computing}} with {{Work Stealing}}},
  author    = {Fernando, Niroshinie and Loke, Seng W. and {W. Rahayu}},
  year      = {2012},
  month     = sep,
  booktitle = {2012 15th {{International Conference}} on {{Network}}-{{Based Information Systems}}},
  publisher = {{IEEE}},
  pages     = {660--665},
  doi       = {10.1109/NBiS.2012.122}
}

@misc{googleOverviewNearbyConnections2018,
  title    = {Overview | {{Nearby Connections API}}},
  author   = {Google},
  year     = {2018},
  journal  = {Google Developers},
  url      = {https://developers.google.com/nearby/connections/overview},
  language = {en}
}

@inproceedings{habakkarimFemtoCloudsLeveraging2015,
  title     = {Femto {{Clouds}}: {{Leveraging Mobile Devices}} to {{Provide Cloud Service}} at the {{Edge}}},
  author    = {Habak, Karim and Ammar, Mostafa and Harras, Khaled A. and Zegura, Ellen},
  year      = {2015},
  month     = jun,
  booktitle = {2015 {{IEEE}} 8th {{International Conference}} on {{Cloud Computing}}},
  publisher = {{IEEE}},
  pages     = {9--16},
  doi       = {10.1109/CLOUD.2015.12}
}

@article{mehrishambujEgocentricAnalysisDashcam2019,
  title    = {Egocentric {{Analysis}} of {{Dash}}-Cam {{Videos For Vehicle Forensics}}},
  author   = {Mehrish, Ambuj and Singh, Prerna and Jain, Puneet and Subramanyam, A. V. and Kankanhalli, Mohan},
  year     = {2019},
  journal  = {IEEE Transactions on Circuits and Systems for Video Technology},
  doi      = {10.1109/TCSVT.2019.2929561},
  issn     = {1558-2205}
}

@inproceedings{parkMotivesConcernsDashcam2016,
  title     = {Motives and Concerns of Dashcam Video Sharing},
  author    = {Park, Sangkeun and Kim, Joohyun and Mizouni, Rabeb and Lee, Uichin},
  year      = {2016},
  booktitle = {Proceedings of the 2016 {{CHI}} Conference on Human Factors in Computing Systems},
  publisher = {{Association for Computing Machinery}},
  address   = {{New York, NY, USA}},
  series    = {{{CHI}} '16},
  pages     = {4758--4769},
  doi       = {10.1145/2858036.2858581},
  isbn      = {978-1-4503-3362-7}
}

@article{mehrishEgocentricAnalysisDashCam2020,
  title   = {Egocentric {{Analysis}} of {{Dash}}-{{Cam Videos}} for {{Vehicle Forensics}}},
  author  = {Mehrish, Ambuj and Singh, Prerna and Jain, Puneet and Subramanyam, A. V. and Kankanhalli, Mohan},
  year    = {2020},
  month   = sep,
  journal = {IEEE Transactions on Circuits and Systems for Video Technology},
  volume  = {30},
  number  = {9},
  pages   = {3000--3014},
  doi     = {10.1109/TCSVT.2019.2929561},
  issn    = {1558-2205}
}

@inproceedings{liMUVRSupportingMultiUser2018,
  title      = {{{MUVR}}: {{Supporting Multi}}-{{User Mobile Virtual Reality}} with {{Resource Constrained Edge Cloud}}},
  shorttitle = {{{MUVR}}},
  author     = {Li, Yong and Gao, Wei},
  year       = {2018},
  month      = oct,
  booktitle  = {2018 {{IEEE}}/{{ACM Symposium}} on {{Edge Computing}} ({{SEC}})},
  pages      = {1--16},
  doi        = {10.1109/SEC.2018.00008}
}

@inproceedings{stamateDeepLearningParkinson2017,
  title     = {Deep Learning {{Parkinson}}'s from Smartphone Data},
  author    = {Stamate, C. and Magoulas, G.D. and Kueppers, S. and Nomikou, E. and Daskalopoulos, I. and Luchini, M.U. and Moussouri, T. and Roussos, G.},
  year      = {2017},
  month     = mar,
  booktitle = {2017 {{IEEE International Conference}} on {{Pervasive Computing}} and {{Communications}} ({{PerCom}})},
  pages     = {31--40},
  doi       = {10.1109/PERCOM.2017.7917848},
  issn      = {2474-249X}
}

@inproceedings{ranDeepDecisionMobileDeep2018,
  title      = {{{DeepDecision}}: {{A Mobile Deep Learning Framework}} for {{Edge Video Analytics}}},
  shorttitle = {{{DeepDecision}}},
  author     = {Ran, Xukan and Chen, Haolianz and Zhu, Xiaodan and Liu, Zhenming and Chen, Jiasi},
  year       = {2018},
  month      = apr,
  booktitle  = {{{IEEE INFOCOM}} 2018 - {{IEEE Conference}} on {{Computer Communications}}},
  pages      = {1421--1429},
  doi        = {10.1109/INFOCOM.2018.8485905}
}

@inproceedings{delahozRealtimeSmartphonebasedFloor2017,
  title     = {A Real-Time Smartphone-Based Floor Detection System for the Visually Impaired},
  author    = {Delahoz, Yueng and Labrador, Miguel A.},
  year      = {2017},
  month     = may,
  booktitle = {2017 {{IEEE International Symposium}} on {{Medical Measurements}} and {{Applications}} ({{MeMeA}})},
  publisher = {{IEEE}},
  pages     = {27--32},
  doi       = {10.1109/MeMeA.2017.7985844}
}

@inproceedings{mandalWearableFaceRecognition2015,
  title     = {A {{Wearable Face Recognition System}} on {{Google Glass}} for {{Assisting Social Interactions}}},
  author    = {Mandal, Bappaditya and Chia, Shue-Ching and Li, Liyuan and Chandrasekhar, Vijay and Tan, Cheston and Lim, Joo-Hwee},
  year      = {2015},
  booktitle = {Computer {{Vision}} - {{ACCV}} 2014 {{Workshops}}},
  publisher = {{Springer International Publishing}},
  address   = {{Cham}},
  series    = {Lecture {{Notes}} in {{Computer Science}}},
  pages     = {419--433},
  doi       = {10.1007/978-3-319-16634-6_31},
  isbn      = {978-3-319-16634-6},
  editor    = {Jawahar, C. V. and Shan, Shiguang}
}

@inproceedings{koukoumidisSignalGuruLeveragingMobile2011,
  title      = {{{SignalGuru}}: Leveraging Mobile Phones for Collaborative Traffic Signal Schedule Advisory},
  shorttitle = {{{SignalGuru}}},
  author     = {Koukoumidis, Emmanouil and Peh, Li-Shiuan and Martonosi, Margaret Rose},
  year       = {2011},
  month      = jun,
  booktitle  = {Proceedings of the 9th International Conference on {{Mobile}} Systems, Applications, and Services},
  publisher  = {{Association for Computing Machinery}},
  address    = {{New York, NY, USA}},
  series     = {{{MobiSys}} '11},
  pages      = {127--140},
  doi        = {10.1145/1999995.2000008},
  isbn       = {978-1-4503-0643-0}
}

@misc{googleTensorFlowLiteTask,
  title   = {{{TensorFlow Lite Task Library}}},
  author  = {{Google}},
  year    = {2021},
  journal = {TensorFlow},
  url     = {https://www.tensorflow.org/lite/inference_with_metadata/task_library/overview},
  urldate = {2021-01-26}
}

@misc{tensorflowEfficientDetLite42021,
  title     = {{{EfficientDet}}-{{Lite4}}},
  author    = {{TensorFlow}},
  year      = {2021},
  journal   = {TensorFlow Hub},
  url       = {https://tfhub.dev/tensorflow/lite-model/efficientdet/lite4/detection/metadata/2},
  urldate   = {2021-07-31},
  copyright = {Apache-2.0 License}
}

@misc{tensorflowSSDMobileNetV12021,
  title     = {{{SSD MobileNet}} V1},
  author    = {{TensorFlow}},
  year      = {2021},
  journal   = {TensorFlow Hub},
  url       = {https://tfhub.dev/tensorflow/lite-model/ssd_mobilenet_v1/1/metadata/2},
  urldate   = {2021-01-26},
  copyright = {Apache-2.0}
}

@misc{tensorflowMoveNetLightning2022,
  title     = {{{MoveNet Lightning}}},
  author    = {TensorFlow},
  year      = {2022},
  journal   = {TensorFlow Hub},
  url       = {https://tfhub.dev/google/lite-model/movenet/singlepose/lightning/tflite/float16/4},
  urldate   = {2022-05-14},
  copyright = {Apache-2.0}
}

@inproceedings{yuBDD100KDiverseDriving2020,
  title         = {{{BDD100K}}: {{A Diverse Driving Dataset}} for {{Heterogeneous Multitask Learning}}},
  shorttitle    = {{{BDD100K}}},
  booktitle     = {2020 {{IEEE}}/{{CVF Conference}} on {{Computer Vision}} and {{Pattern Recognition}} ({{CVPR}})},
  author        = {Yu, Fisher and Chen, Haofeng and Wang, Xin and Xian, Wenqi and Chen, Yingying and Liu, Fangchen and Madhavan, Vashisht and Darrell, Trevor},
  year          = {2020},
  month         = jun,
  eprint        = {1805.04687},
  eprinttype    = {arxiv},
  primaryclass  = {cs.CV},
  pages         = {2633--2642},
  issn          = {2575-7075},
  doi           = {10.1109/CVPR42600.2020.00271},
  archiveprefix = {arXiv}
}

@article{silvaEnergyawareAdaptiveOffloading2021,
  title         = {Energy-Aware Adaptive Offloading of Soft Real-Time Jobs in Mobile Edge Clouds},
  author        = {Silva, Joaquim and Marques, Eduardo R. B. and Lopes, Lu{\'i}s M.B. and Silva, Fernando},
  year          = {2021},
  month         = jul,
  journal       = {Journal of Cloud Computing},
  volume        = {10},
  number        = {38},
  eprint        = {2102.05504},
  eprinttype    = {arxiv},
  primaryclass  = {cs.DC},
  pages         = {1--21},
  issn          = {2192-113X},
  doi           = {10.1186/s13677-021-00251-9},
  archiveprefix = {arXiv}
}

@inproceedings{ortegaDMDLargeScaleMultimodal2020,
  title         = {{{DMD}}: {{A Large-Scale Multi-modal Driver Monitoring Dataset}} for {{Attention}} and {{Alertness Analysis}}},
  shorttitle    = {{{DMD}}},
  booktitle     = {Computer {{Vision}} \textendash{} {{ECCV}} 2020 {{Workshops}}},
  author        = {Ortega, Juan Diego and Kose, Neslihan and Ca{\~n}as, Paola and Chao, Min-An and Unnervik, Alexander and Nieto, Marcos and Otaegui, Oihana and Salgado, Luis},
  editor        = {Bartoli, Adrien and Fusiello, Andrea},
  year          = {2020},
  series        = {Lecture {{Notes}} in {{Computer Science}}},
  eprint        = {2008.12085},
  eprinttype    = {arxiv},
  primaryclass  = {cs.CV},
  pages         = {387--405},
  publisher     = {{Springer International Publishing}},
  address       = {{Cham}},
  doi           = {10.1007/978-3-030-66823-5_23},
  archiveprefix = {arXiv},
  isbn          = {978-3-030-66823-5}
}

@misc{thetensorflowauthorsTensorFlowLiteSupport2021,
  title        = {{{TensorFlow Lite Support}}},
  shorttitle   = {{{TFLite Support}}},
  author       = {{The TensorFlow Authors}},
  year         = {2021},
  month        = jan,
  url          = {https://github.com/tensorflow/tflite-support},
  urldate      = {2021-01-26},
  copyright    = {Apache-2.0 License},
  howpublished = {GitHub repository}
}

@misc{francisFetch2022,
  title     = {Fetch},
  author    = {Francis, Tonyo},
  year      = {2022},
  month     = feb,
  url       = {https://github.com/tonyofrancis/Fetch},
  urldate   = {2022-02-18},
  copyright = {Apache-2.0}
}

@misc{hedleyJsoup2022,
  title      = {jsoup},
  shorttitle = {jsoup},
  author     = {Hedley, Jonathan},
  year       = {2022},
  month      = may,
  url        = {https://github.com/jhy/jsoup},
  urldate    = {2022-05-15},
  copyright  = {MIT}
}

@misc{googlePixelPhoneHardware2022,
  title   = {Pixel Phone Hardware Tech Specs},
  author  = {{Google}},
  year    = {2022},
  journal = {Pixel Phone Help},
  url     = {https://support.google.com/pixelphone/answer/7158570},
  urldate = {2022-05-19}
}

@misc{kressin2019Snapdragon8652019,
  title      = {2019 {{Snapdragon}} 865 {{Deep Dives Intro}}},
  shorttitle = {2019 {{Snapdragon}} 865 {{5G Deep Dives Intro}}},
  author     = {Kressin, Keith and Patrick, Chris and Seed, Jesse},
  year       = {2019},
  month      = dec,
  journal    = {Qualcomm},
  url        = {https://www.qualcomm.com/content/dam/qcomm-martech/dm-assets/documents/2019_snapdragon_865_deep_dive_-_intro_-_keith_kressin_-_chris_patrick_-_jesse_seed.pdf},
  urldate    = {2022-05-19}
}

@misc{oneplusOnePlusSpecs2020,
  title   = {{{OnePlus}} 8 {{Specs}}},
  author  = {{OnePlus}},
  year    = {2020},
  journal = {OnePlus},
  url     = {https://web.archive.org/web/20200421084636/https://www.oneplus.com/8/specs},
  urldate = {2022-05-19}
}

@misc{oppoOPPOFindX22022,
  title   = {{OPPO Find X2 Pro Specifications}},
  author  = {{OPPO}},
  year    = {2022},
  journal = {OPPO},
  url     = {https://www.oppo.com/en/smartphone-find-x2-pro/specs/},
  urldate = {2022-05-19}
}

@misc{tibkenGoogleBuiltAI2021,
  title   = {Google Built an {{AI}} Chip to Make the {{Pixel}} 6 Smarter and Last Longer},
  author  = {Tibken, Shara},
  year    = {2021},
  month   = oct,
  journal = {CNET},
  url     = {https://www.cnet.com/tech/mobile/pixel-6s-tensor-chip-inside-the-brains-of-googles-newest-flagship/},
  urldate = {2022-05-19},
  langid  = {english}
}

@article{dingusDriverCrashRisk2016,
  title     = {Driver Crash Risk Factors and Prevalence Evaluation Using Naturalistic Driving Data},
  author    = {Dingus, Thomas A. and Guo, Feng and Lee, Suzie and Antin, Jonathan F. and Perez, Miguel and {Buchanan-King}, Mindy and Hankey, Jonathan},
  year      = {2016},
  month     = mar,
  journal   = {Proceedings of the National Academy of Sciences},
  volume    = {113},
  number    = {10},
  pages     = {2636--2641},
  publisher = {{Proceedings of the National Academy of Sciences}},
  doi       = {10.1073/pnas.1513271113}
}

@inproceedings{rezaeiLookDriverLook2014,
  title      = {Look at the {{Driver}}, {{Look}} at the {{Road}}: {{No Distraction}}! {{No Accident}}!},
  shorttitle = {Look at the {{Driver}}, {{Look}} at the {{Road}}},
  booktitle  = {2014 {{IEEE Conference}} on {{Computer Vision}} and {{Pattern Recognition}}},
  author     = {Rezaei, Mahdi and Klette, Reinhard},
  year       = {2014},
  month      = jun,
  pages      = {129--136},
  issn       = {1063-6919},
  doi        = {10.1109/CVPR.2014.24}
}

@inproceedings{jainCarThatKnows2015,
  title         = {Car That {{Knows Before You Do}}: {{Anticipating Maneuvers}} via {{Learning Temporal Driving Models}}},
  shorttitle    = {Car That {{Knows Before You Do}}},
  booktitle     = {2015 {{IEEE International Conference}} on {{Computer Vision}} ({{ICCV}})},
  author        = {Jain, Ashesh and Koppula, Hema S. and Raghavan, Bharad and Soh, Shane and Saxena, Ashutosh},
  year          = {2015},
  month         = dec,
  eprint        = {1504.02789},
  eprinttype    = {arxiv},
  primaryclass  = {cs.CV},
  pages         = {3182--3190},
  issn          = {2380-7504},
  doi           = {10.1109/ICCV.2015.364},
  archiveprefix = {arXiv}
}

@article{trivediLookingInLookingOutVehicle2007,
  title      = {Looking-{{In}} and {{Looking-Out}} of a {{Vehicle}}: {{Computer-Vision-Based Enhanced Vehicle Safety}}},
  shorttitle = {Looking-{{In}} and {{Looking-Out}} of a {{Vehicle}}},
  author     = {Trivedi, Mohan Manubhai and Gandhi, Tarak and McCall, Joel},
  year       = {2007},
  month      = mar,
  journal    = {IEEE Transactions on Intelligent Transportation Systems},
  volume     = {8},
  number     = {1},
  pages      = {108--120},
  issn       = {1558-0016},
  doi        = {10.1109/TITS.2006.889442}
}

@article{leeRealTimeGazeEstimator2011,
  title   = {Real-{{Time Gaze Estimator Based}} on {{Driver}}'s {{Head Orientation}} for {{Forward Collision Warning System}}},
  author  = {Lee, Sung Joo and Jo, Jaeik and Jung, Ho Gi and Park, Kang Ryoung and Kim, Jaihie},
  year    = {2011},
  month   = mar,
  journal = {IEEE Transactions on Intelligent Transportation Systems},
  volume  = {12},
  number  = {1},
  pages   = {254--267},
  issn    = {1558-0016},
  doi     = {10.1109/TITS.2010.2091503}
}

@inproceedings{choiRealtimeCategorizationDriver2016,
  title     = {Real-Time Categorization of Driver's Gaze Zone Using the Deep Learning Techniques},
  booktitle = {2016 {{International Conference}} on {{Big Data}} and {{Smart Computing}} ({{BigComp}})},
  author    = {Choi, In-Ho and Hong, Sung Kyung and Kim, Yong-Guk},
  year      = {2016},
  month     = jan,
  pages     = {143--148},
  issn      = {2375-9356},
  doi       = {10.1109/BIGCOMP.2016.7425813}
}

@article{voraDriverGazeZone2018,
  title         = {Driver {{Gaze Zone Estimation Using Convolutional Neural Networks}}: {{A General Framework}} and {{Ablative Analysis}}},
  shorttitle    = {Driver {{Gaze Zone Estimation Using Convolutional Neural Networks}}},
  author        = {Vora, Sourabh and Rangesh, Akshay and Trivedi, Mohan Manubhai},
  year          = {2018},
  month         = sep,
  journal       = {IEEE Transactions on Intelligent Vehicles},
  volume        = {3},
  number        = {3},
  eprint        = {1802.02690},
  eprinttype    = {arxiv},
  primaryclass  = {cs.CV},
  pages         = {254--265},
  issn          = {2379-8904},
  doi           = {10.1109/TIV.2018.2843120},
  archiveprefix = {arXiv}
}

@inproceedings{yanDriverBehaviorRecognition2016,
  title     = {Driver Behavior Recognition Based on Deep Convolutional Neural Networks},
  booktitle = {2016 12th {{International Conference}} on {{Natural Computation}}, {{Fuzzy Systems}} and {{Knowledge Discovery}} ({{ICNC-FSKD}})},
  author    = {Yan, Shiyang and Teng, Yuxuan and Smith, Jeremy S. and Zhang, Bailing},
  year      = {2016},
  month     = aug,
  pages     = {636--641},
  doi       = {10.1109/FSKD.2016.7603248}
}

@article{tranRealtimeDetectionDistracted2018,
  title   = {Real-Time Detection of Distracted Driving Based on Deep Learning},
  author  = {Tran, Duy and Manh Do, Ha and Sheng, Weihua and Bai, He and Chowdhary, Girish},
  year    = {2018},
  journal = {IET Intelligent Transport Systems},
  volume  = {12},
  number  = {10},
  pages   = {1210--1219},
  issn    = {1751-9578},
  doi     = {10.1049/iet-its.2018.5172}
}

@inproceedings{kapoorRealTimeDriverDistraction2020,
  title     = {Real-{{Time Driver Distraction Detection System Using Convolutional Neural Networks}}},
  booktitle = {Proceedings of {{ICETIT}} 2019},
  author    = {Kapoor, Khyati and Pamula, Rajendra and Murthy, Sristi Vns},
  editor    = {Singh, Pradeep Kumar and Panigrahi, Bijaya Ketan and Suryadevara, Nagender Kumar and Sharma, Sudhir Kumar and Singh, Amit Prakash},
  year      = {2020},
  series    = {Lecture {{Notes}} in {{Electrical Engineering}}},
  pages     = {280--291},
  publisher = {{Springer International Publishing}},
  address   = {{Cham}},
  doi       = {10.1007/978-3-030-30577-2_24},
  isbn      = {978-3-030-30577-2}
}

@article{huangHCFHybridCNN2020,
  title      = {{{HCF}}: {{A Hybrid CNN Framework}} for {{Behavior Detection}} of {{Distracted Drivers}}},
  shorttitle = {{{HCF}}},
  author     = {Huang, Chen and Wang, Xiaochen and Cao, Jiannong and Wang, Shihui and Zhang, Yan},
  year       = {2020},
  journal    = {IEEE Access},
  volume     = {8},
  pages      = {109335--109349},
  issn       = {2169-3536},
  doi        = {10.1109/ACCESS.2020.3001159}
}

@inproceedings{creusotRealtimeSmallObstacle2015,
  title     = {Real-Time Small Obstacle Detection on Highways Using Compressive {{RBM}} Road Reconstruction},
  booktitle = {2015 {{IEEE Intelligent Vehicles Symposium}} ({{IV}})},
  author    = {Creusot, Clement and Munawar, Asim},
  year      = {2015},
  month     = jun,
  pages     = {162--167},
  issn      = {1931-0587},
  doi       = {10.1109/IVS.2015.7225680}
}

@article{moralesrosalesOnroadObstacleDetection2018,
  title     = {On-Road Obstacle Detection Video System for Traffic Accident Prevention},
  author    = {Morales Rosales, Luis Alberto and Algredo Badillo, Ignacio and Hern{\'a}ndez Gracidas, Carlos Arturo and Rangel, Hector Rodr{\'i}guez and Lobato B{\'a}ez, Mariana and Patnaik, Srikanta},
  year      = {2018},
  month     = jul,
  journal   = {Journal of Intelligent \& Fuzzy Systems},
  volume    = {35},
  number    = {1},
  pages     = {533--547},
  publisher = {{IOS Press}},
  issn      = {10641246},
  doi       = {10.3233/JIFS-169609}
}

@article{jiaRealtimeObstacleDetection2015,
  title   = {Real-Time Obstacle Detection with Motion Features Using Monocular Vision},
  author  = {Jia, Baozhi and Liu, Rui and Zhu, Ming},
  year    = {2015},
  month   = mar,
  journal = {The Visual Computer},
  volume  = {31},
  number  = {3},
  pages   = {281--293},
  issn    = {1432-2315},
  doi     = {10.1007/s00371-014-0918-5}
}

@article{ahmedPedestrianCyclistDetection2019,
  title      = {Pedestrian and {{Cyclist Detection}} and {{Intent Estimation}} for {{Autonomous Vehicles}}: {{A Survey}}},
  shorttitle = {Pedestrian and {{Cyclist Detection}} and {{Intent Estimation}} for {{Autonomous Vehicles}}},
  author     = {Ahmed, Sarfraz and Huda, M. Nazmul and Rajbhandari, Sujan and Saha, Chitta and Elshaw, Mark and Kanarachos, Stratis},
  year       = {2019},
  month      = jan,
  journal    = {Applied Sciences},
  volume     = {9},
  number     = {11},
  pages      = {2335},
  publisher  = {{Multidisciplinary Digital Publishing Institute}},
  issn       = {2076-3417},
  doi        = {10.3390/app9112335},
  copyright  = {CC BY 3.0}
}

@article{brunettiComputerVisionDeep2018,
  title      = {Computer Vision and Deep Learning Techniques for Pedestrian Detection and Tracking: {{A}} Survey},
  shorttitle = {Computer Vision and Deep Learning Techniques for Pedestrian Detection and Tracking},
  author     = {Brunetti, Antonio and Buongiorno, Domenico and Trotta, Gianpaolo Francesco and Bevilacqua, Vitoantonio},
  year       = {2018},
  month      = jul,
  journal    = {Neurocomputing},
  volume     = {300},
  pages      = {17--33},
  issn       = {0925-2312},
  doi        = {10.1016/j.neucom.2018.01.092}
}

@article{hurneyReviewPedestrianDetection2015,
  title   = {Review of Pedestrian Detection Techniques in Automotive Far-Infrared Video},
  author  = {Hurney, Patrick and Waldron, Peter and Morgan, Fearghal and Jones, Edward and Glavin, Martin},
  year    = {2015},
  journal = {IET Intelligent Transport Systems},
  volume  = {9},
  number  = {8},
  pages   = {824--832},
  issn    = {1751-9578},
  doi     = {10.1049/iet-its.2014.0236}
}

@article{mukhtarVehicleDetectionTechniques2015,
  title      = {Vehicle {{Detection Techniques}} for {{Collision Avoidance Systems}}: {{A Review}}},
  shorttitle = {Vehicle {{Detection Techniques}} for {{Collision Avoidance Systems}}},
  author     = {Mukhtar, Amir and Xia, Likun and Tang, Tong Boon},
  year       = {2015},
  month      = oct,
  journal    = {IEEE Transactions on Intelligent Transportation Systems},
  volume     = {16},
  number     = {5},
  pages      = {2318--2338},
  issn       = {1558-0016},
  doi        = {10.1109/TITS.2015.2409109}
}

@article{sivaramanLookingVehiclesRoad2013,
  title      = {Looking at {{Vehicles}} on the {{Road}}: {{A Survey}} of {{Vision-Based Vehicle Detection}}, {{Tracking}}, and {{Behavior Analysis}}},
  shorttitle = {Looking at {{Vehicles}} on the {{Road}}},
  author     = {Sivaraman, Sayanan and Trivedi, Mohan Manubhai},
  year       = {2013},
  month      = dec,
  journal    = {IEEE Transactions on Intelligent Transportation Systems},
  volume     = {14},
  number     = {4},
  pages      = {1773--1795},
  issn       = {1558-0016},
  doi        = {10.1109/TITS.2013.2266661}
}

@article{sunOnroadVehicleDetection2006,
  title      = {On-Road Vehicle Detection: A Review},
  shorttitle = {On-Road Vehicle Detection},
  author     = {Sun, Zehang and Bebis, G. and Miller, R.},
  year       = {2006},
  month      = may,
  journal    = {IEEE Transactions on Pattern Analysis and Machine Intelligence},
  volume     = {28},
  number     = {5},
  pages      = {694--711},
  issn       = {1939-3539},
  doi        = {10.1109/TPAMI.2006.104}
}

@inproceedings{chadwickDistantVehicleDetection2019,
  title         = {Distant {{Vehicle Detection Using Radar}} and {{Vision}}},
  booktitle     = {2019 {{International Conference}} on {{Robotics}} and {{Automation}} ({{ICRA}})},
  author        = {Chadwick, Simon and Maddern, Will and Newman, Paul},
  year          = {2019},
  month         = may,
  eprint        = {1901.10951},
  eprinttype    = {arxiv},
  primaryclass  = {cs.RO},
  pages         = {8311--8317},
  issn          = {2577-087X},
  doi           = {10.1109/ICRA.2019.8794312},
  archiveprefix = {arXiv}
}

@inproceedings{rybskiVisualClassificationCoarse2010,
  title     = {Visual Classification of Coarse Vehicle Orientation Using {{Histogram}} of {{Oriented Gradients}} Features},
  booktitle = {2010 {{IEEE Intelligent Vehicles Symposium}}},
  author    = {Rybski, Paul E. and Huber, Daniel and Morris, Daniel D. and Hoffman, Regis},
  year      = {2010},
  month     = jun,
  pages     = {921--928},
  issn      = {1931-0587},
  doi       = {10.1109/IVS.2010.5547996}
}

@article{tehraniniknejadOnRoadMultivehicleTracking2012,
  title   = {On-{{Road Multivehicle Tracking Using Deformable Object Model}} and {{Particle Filter With Improved Likelihood Estimation}}},
  author  = {Tehrani Niknejad, Hossein and Takeuchi, Akihiro and Mita, Seiichi and McAllester, David},
  year    = {2012},
  month   = jun,
  journal = {IEEE Transactions on Intelligent Transportation Systems},
  volume  = {13},
  number  = {2},
  pages   = {748--758},
  issn    = {1558-0016},
  doi     = {10.1109/TITS.2012.2187894}
}

@article{zhangRealtimeVehicleDetection2018,
  title     = {Real-Time Vehicle Detection and Tracking Using Improved Histogram of Gradient Features and {{Kalman}} Filters},
  author    = {Zhang, Xinyu and Gao, Hongbo and Xue, Chong and Zhao, Jianhui and Liu, Yuchao},
  year      = {2018},
  month     = jan,
  journal   = {International Journal of Advanced Robotic Systems},
  volume    = {15},
  number    = {1},
  pages     = {1729881417749949},
  publisher = {{SAGE Publications}},
  issn      = {1729-8806},
  doi       = {10.1177/1729881417749949}
}

@article{dominguez-sanchezPedestrianMovementDirection2017,
  title   = {Pedestrian {{Movement Direction Recognition Using Convolutional Neural Networks}}},
  author  = {{Dominguez-Sanchez}, Alex and Cazorla, Miguel and {Orts-Escolano}, Sergio},
  year    = {2017},
  month   = dec,
  journal = {IEEE Transactions on Intelligent Transportation Systems},
  volume  = {18},
  number  = {12},
  pages   = {3540--3548},
  issn    = {1558-0016},
  doi     = {10.1109/TITS.2017.2726140}
}

@article{tomeDeepConvolutionalNeural2016,
  title         = {Deep {{Convolutional Neural Networks}} for Pedestrian Detection},
  author        = {Tom{\`e}, D. and Monti, F. and Baroffio, L. and Bondi, L. and Tagliasacchi, M. and Tubaro, S.},
  year          = {2016},
  month         = sep,
  journal       = {Signal Processing: Image Communication},
  volume        = {47},
  eprint        = {1510.03608},
  eprinttype    = {arxiv},
  primaryclass  = {cs.CV},
  pages         = {482--489},
  issn          = {0923-5965},
  doi           = {10.1016/j.image.2016.05.007},
  archiveprefix = {arXiv}
}

@article{liuRobustFastPedestrian2013,
  title   = {Robust and Fast Pedestrian Detection Method for Far-Infrared Automotive Driving Assistance Systems},
  author  = {Liu, Qiong and Zhuang, Jiajun and Ma, Jun},
  year    = {2013},
  month   = sep,
  journal = {Infrared Physics \& Technology},
  volume  = {60},
  pages   = {288--299},
  issn    = {1350-4495},
  doi     = {10.1016/j.infrared.2013.06.003}
}

@article{brostowSemanticObjectClasses2009,
  title      = {Semantic Object Classes in Video: {{A}} High-Definition Ground Truth Database},
  shorttitle = {Semantic Object Classes in Video},
  author     = {Brostow, Gabriel J. and Fauqueur, Julien and Cipolla, Roberto},
  year       = {2009},
  month      = jan,
  journal    = {Pattern Recognition Letters},
  series     = {Video-Based {{Object}} and {{Event Analysis}}},
  volume     = {30},
  number     = {2},
  pages      = {88--97},
  issn       = {0167-8655},
  doi        = {10.1016/j.patrec.2008.04.005}
}

@inproceedings{geigerAreWeReady2012,
  title      = {Are We Ready for Autonomous Driving? {{The KITTI}} Vision Benchmark Suite},
  shorttitle = {Are We Ready for Autonomous Driving?},
  booktitle  = {2012 {{IEEE Conference}} on {{Computer Vision}} and {{Pattern Recognition}}},
  author     = {Geiger, Andreas and Lenz, Philip and Urtasun, Raquel},
  year       = {2012},
  month      = jun,
  pages      = {3354--3361},
  issn       = {1063-6919},
  doi        = {10.1109/CVPR.2012.6248074}
}

@misc{googleTensorFlowLiteML2021,
  title   = {{{TensorFlow Lite}} | {{ML}} for {{Mobile}} and {{Edge Devices}}},
  author  = {{Google}},
  year    = {2021},
  journal = {TensorFlow},
  url     = {https://www.tensorflow.org/lite},
  urldate = {2021-10-18}
}

@misc{googleNearbyConnections2021,
  title        = {Nearby {{Connections}}},
  author       = {{Google}},
  year         = {2021},
  url          = {https://github.com/google/nearby},
  urldate      = {2020-07-15},
  copyright    = {Apache-2.0},
  howpublished = {Google}
}

@misc{kozlenkoWombatCrossingRoad2012,
  title      = {Wombat Crossing Road, near {{Thredbo}}},
  shorttitle = {English},
  author     = {Kozlenko, Maksym},
  year       = {2012},
  month      = may,
  url        = {https://commons.wikimedia.org/wiki/File:Wombat_crossing_road.jpg},
  urldate    = {2022-06-17},
  copyright  = {CC BY-SA 4.0}
}

\end{document}